\documentclass{nature}
\usepackage{graphicx} 
\usepackage{subfig}
\usepackage[dvipsnames]{xcolor}
\usepackage{amsmath}
\usepackage{mathrsfs}
\usepackage{bm}
\usepackage{qcircuit}
\usepackage{hyperref}
\usepackage{leftidx}
\usepackage{enumerate}
\usepackage{times}
\usepackage{float}
\usepackage{amsmath}
\usepackage{amssymb}
\newcommand{\ket}[1]{|#1\rangle}

\makeatletter
\let\saved@includegraphics\includegraphics
\AtBeginDocument{\let\includegraphics\saved@includegraphics}
\renewenvironment*{figure}{\@float{figure}}{\end@float}
\makeatother

\usepackage{dsfont}
\usepackage{widetext}
\usepackage[super,numbers]{natbib}

\title{Testing a Quantum Error-Correcting Code on Various Platforms}

\author{Qihao Guo$^{1,2,3,*}$, Yuan-Yuan Zhao$^{4,3,*}$, Markus Grassl$^{5,6} $, Xinfang Nie$ ^{7,3} $, Guo-Yong Xiang$^4 $, Tao Xin$ ^{7,3} $, Zhang-Qi Yin$^8 $, \& Bei Zeng$^{9,10,11}$}

\begin{document}
\spacing{1}
\maketitle

\begin{affiliations}
 \item Institute for Quantum Computing, Baidu Research, Beijing 100193, China
 \item Department of Applied Physics, Xi’an Jiaotong University, Xi’an, Shaanxi 710049, China
 \item Center for Quantum Computing, Peng Cheng Laboratory, Shenzhen 518055, China
 \item CAS Key Laboratory of Quantum Information, University of Science and Technology of China, Hefei 230026, China
 \item Max Planck Institute for the Science of Light, 91058 Erlangen, Germany
 \item International Centre for Theory of Quantum Technologies, 80-308 Gda\'nsk, Poland
 \item Shenzhen Institute for Quantum Science and Engineering, Southern University of Science and Technology, Shenzhen 518055, China
 \item Center for Quantum Technology Research, School of Physics, Beijing Institute of Technology, Beijing 100081, China
 \item Department of Physics, The Hong Kong University of Science and Technology, Clear Water Bay, Kowloon, Hong Kong
 \item Department of Mathematics $\&$ Statistics, University of Guelph, Guelph, Ontario, N1G 2W1, Canada
 \item Institute for Quantum Computing, University of Waterloo, Waterloo, Ontario, N2L 3G1, Canada
\end{affiliations}

\begin{abstract}
Quantum error correction plays an important role in fault-tolerant
quantum information processing.  It is usually difficult to
experimentally realize quantum error correction, as it requires
multiple qubits and quantum gates with high fidelity.  Here we propose
a simple quantum error-correcting code for the detected amplitude
damping channel. The code requires only two qubits. We implement the
encoding, the channel, and the recovery on an optical platform, the
IBM Q System, and a nuclear magnetic resonance system. 
For all of these systems, the
error correction advantage appears when the damping rate exceeds some
threshold.  We compare the features of these quantum information
processing systems used and demonstrate the advantage of quantum
error correction on current quantum computing platforms.
\end{abstract}

\section{Introduction}

Quantum computing, as the next generation of information technology,
exploits the superposition principle and quantum entanglement to solve
some difficult problems more efficiently than classical computing
devices.  It is widely believed that quantum computing has potential
to realize an exponential advantage for certain problems, such as
prime factor decomposition\cite{shor1994algorithms} and principal
component analysis\cite{lloyd2014quantum}, over current classical
algorithms.  In addition, some pioneering work also connects quantum
computing with other research fields, including quantum simulation,
cryptography, and machine learning.  Since the concept of quantum
computers came into being, several quantum systems, such as linear
optical systems, nuclear magnetic resonance (NMR) systems, trapped ion
systems, and superconducting circuits, were regarded as possible
platforms to implement quantum computers\cite{ladd2010quantum}.  Over
the past decade, hardware for quantum computers has undergone an
astonishing evolution, especially on superconducting circuits and
trapped ion systems. Very recently, Google announced that they had achieved
quantum advantage using a programmable superconducting processor with
$53$ qubits\cite{arute2019quantum}.  In the field of trapped ions,
IonQ also made a presentation about their quantum computer with $79$
processing qubits\cite{ionQ}.  On the other hand, IBM and Rigetti released their
online quantum platforms linking with real superconducting quantum
computers to the public.  We are now entering a new era in quantum
technology, namely the Noisy Intermediate-Scale Quantum
(NISQ)\cite{preskill2018quantum} era, even with fault-tolerant quantum
computing still a distant dream.

Theoretically, quantum computers could outperform classical computers
dramatically. However, it still presents a major obstacle that the
information encoded on qubits is very vulnerable to the noise induced
by inevitable interaction between the qubits and the environment.
Almost all the proposed physical implementations encounter quantum
errors, including decoherence, imperfect quantum logic gates, and
readout error.  A direct approach to reduce quantum errors is
improving the quantum computers on the physical level. At present, in
superconducting quantum processors, single-qubit and two-qubit gate
fidelities exceed $99.9~\%$ and $99.5~\%$\cite{arute2019quantum},
respectively.  Benefiting from well-developed quantum control techniques, such as composite pulses\cite{levitt1986composite}, refocusing pulses\cite{bendall1983depth}, and the Gradient Ascent Pulse
Engineering (GRAPE) algorithm\cite{rowland2012implementing}, fidelities of quantum
gates can reach even higher accuracy on NMR quantum computers.

While improving the quantum hardware is in the main focus of research
right now, it is impossible to completely eliminate the errors in
quantum computers.  To realize a reliable quantum computer, additional
techniques are required.  Quantum error correction (QEC)
\cite{calderbank1996good,gottesman2002introduction,Nielsen:2011:QCQ:1972505},
protecting quantum information against unwanted operations, has
spawned considerable interest from both physicists and
mathematicians. Some initial theoretical results in this field focused
on quantum error-correcting codes (QECC)
\cite{leung1997approximate,chao2018quantum,grassl2018quantum,beale2018quantum,divincenzo1996fault},
other approaches are noiseless quantum codes and decoherence free
subspaces\cite{PhysRevLett.82.4556}. The discovery of QECCs enhanced the possibility of
building a quantum computer and has further lead to the concept of
fault-tolerant quantum computation
\cite{divincenzo1996fault,divincenzo2000physical,steane2003overhead}. One
important QECC is the surface code
with a fault tolerance threshold of $1 \times 10^{-2}$ for each error
source\cite{wang2011surface,barends2014logic}.  Previous experimental
progress for some quantum error-correcting codes demonstrated the
power of QECC for several qubits for linear optics
\cite{braunstein1998quantum}, trapped ions
\cite{zhang2019error,schindler2011experimental}, NMR
\cite{cory1998experimental}, and superconducting circuits
\cite{ofek2016extending,reed2012realization,rosenblum2018fault,hu2019quantum}. Measurement-based
feedback \cite{hu2019quantum,cardona2019continuous} and other advanced
techniques have also been developed to implement error correction, in
order to build a continuous-time and automatic quantum error
correction system.

\begin{figure}
	\centering
	\includegraphics[width=6.0in]{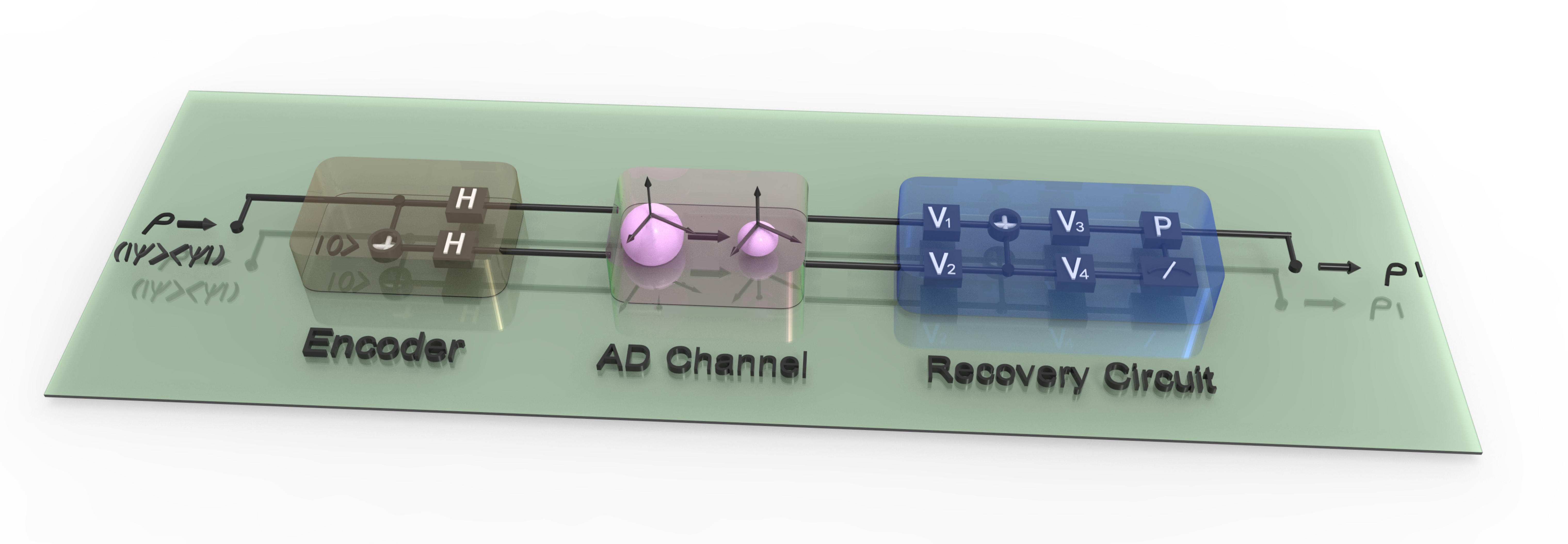}
	\caption{\label{Fig.Model}The model of the quantum communication system with an amplitude damping channel. 
		  The encoder maps an arbitrary initial
          single-qubit state $\rho=|\psi\rangle \langle \psi|$ to the
          code space using two qubits. Then a detected amplitude
          damping channel acts on each of the two qubits.  Finally we
          apply the recovery circuit (including decoding), which discards
          the second qubit, obtaining a single-qubit output state
          $\rho'$ that ideally has a large overlap with the input state $|\psi\rangle$.}
\end{figure}

In this paper, we report on the implementation of a channel-adapted
detected amplitude quantum code using a two-qubit system on
various platforms: a quantum optical system, the IBM Q
Experience superconducting circuit, and an NMR quantum system.  The
experiments on different quantum systems successfully demonstrate the
power of the error-correcting code with observable improvement of the
fidelity when the damping rate is larger than a threshold $\gamma_c$.
	
\section{Result}
\subsection{QECC for Detected Amplitude Channel}
In a typical quantum information process, like the one shown in
Fig.~1, quantum information might be subject to spontaneous decay with
detected photon emission, which is modelled by the dectected amplitude
channel.  Generally, a dectected amplitude damping channel is composed
of an amplitude damping channel (denoted by $ \Phi_{AD} $, see the
Supplementary Material) and an ancilla system indicating whether
damping has ocurred. The channel can be
described by Kraus operators with an extra qubit,
\begin{equation}\label{key}
\Phi_{DJ}(\rho)=\sum_{i} \big(A_i \rho A_i^\dagger \big) \otimes |i\rangle\langle i|_{anc}\,,
\end{equation}
where $
A_0=\begin{pmatrix} 1   & 0 \\ 0 & \sqrt{1-\gamma} \end{pmatrix} \ \text{and} \ A_1=\begin{pmatrix} 0   &  \sqrt{\gamma} \\ 0 & 0 \end{pmatrix} \ .
$

The construction of quantum error-correcting codes for the detected
amplitude channel has been discussed in Refs.
\citenum{grassl2010quantum,alber2001stabilizing,alber2003detected}. The
simplest code correcting a single error of the detected amplitude
channel needs only two qubits, and hence can be implemented on a present
quantum computer.  Based on the analysis in Ref. \citenum{grassl2010quantum},
we firstly encode the initial state
$|\psi\rangle=\alpha|0\rangle+\beta|1\rangle$ onto the basis
$|+\rangle|+\rangle$ and $|-\rangle|-\rangle$ using a CNOT gate
followed by two Hadamard gates,
\begin{equation}\label{Eq.2}
\alpha|0\rangle +\beta|1\rangle \to \alpha|+\rangle|+\rangle+\beta|-\rangle|-\rangle.
\end{equation}

For the two-qubit code given by Eq.~\ref{Eq.2}, there are two standard
error correction protocols derived from the classical code, denoted by
Standard A and Standard B.  Additionally, using the polar
decomposition method in the Supplementary Material, we obtain Optimal
Recovery which will lead a higher fidelity.  The optimal recovery
operation will depend on the damping parameter $\gamma$, meaning it is
a ``channel-adaptive" error correction scheme.  The detailed process
for the different protocols is given in the Methods section.

\subsection{Physical Systems}
Photons as a kind of ``flying qubits'' are widely used for quantum
information processing and simulation.  In a linear optics system,
single qubit operations can be implemented with high fidelity as
photons are essentially decoherence-free and are not affected by the
environment.  However, two-qubit gates, like the controlled-NOT (CNOT)
gate, become a challenge as it is difficult to let photons interact.
We are using optical qubits encoded in the polarization degree
of freedom to demonstrate quantum error correction.

In 2018, IBM Q released a $14$-qubit transmon superconducting quantum
processor (Fig.~\ref{Fig.Device}~(b)), \textit{IBM Q 16 Melbourne},
which is accessible via Qiskit, an open-source framework for quantum
computing on IBM Q Experience. The average fidelity of single qubit
operations exceeds $99.0~\%$, and the fidelity of the CNOT operation
is nearly $82.7~\%$ to $95.2~\%$. The pulse durations are $100$~ns and
$348$~ns for single qubit rotation gates and CNOT gates based on the
cross-resonance interaction, respectively.  In addition, two-qubit
gates are only permitted between neighboring qubits that are connected
by a superconducting bus resonator (see the inset in
Fig.~\ref{Fig.Device}~(b)). More information on the qubits and quantum gates
on \textit{IBM Q 16 Melbourne}, such as the dephasing times and gate
fidelities, can be found on the IBM Q site
\url{https://quantumexperience.ng.bluemix.net/qx/devices}.

Nuclear magnetic resonance (NMR) quantum computing is one of the first
proposed schemes for building a quantum computer with spin-$1/2$
nuclei, such as $\leftidx{^{1}}{}\mathrm{H} $ or $ \leftidx{^{13}}{}\mathrm{C} $. With a time-varying radio frequency (RF) field and the free evolution between the different spins, arbitrary unitary transforms can be
implemented in the NMR quantum computer. In our experiment, we used a
Crotonic acid specimen. The four qubits on the Crotonic acid are
represented by the spin-1/2 $\leftidx{^{13}}{}\mathrm{C}$ nuclear spins, labeled as
$\mathrm{C}_1$ to $\mathrm{C}_4$ as shown in Fig.~\ref{Fig.Device}~(c). The decoherence
times of the Crotonic acid are $T_1\approx 1500$~ms and
$T_2^*\approx550$~ms. All NMR experiments were carried out on a Bruker
DRX 600~MHZ spectrometer at room temperature.

\subsection{Experimental Schemes}
For the three quantum systems, quantum optical platform, IBM Q
superconducting circuit and NMR system (see the Supplementary
Material), we have implemented different variants of quantum
error-correction for the detected amplitude damping channel. In this model
of decoherence, an excited state decays to the ground state with some
probability. Monitoring the system, one obtains the addtional
classical information whether the system has decayed or not.  Owing to
the features of the different systems, we first adapt our scheme to
the particular device and decompose the quantum circuits into basic
gates native for each system. In Fig.~3, we give the quantum circuits
that we employed in the realistic experimental process.

\begin{figure}
	\centering
	\includegraphics[width=6.0in]{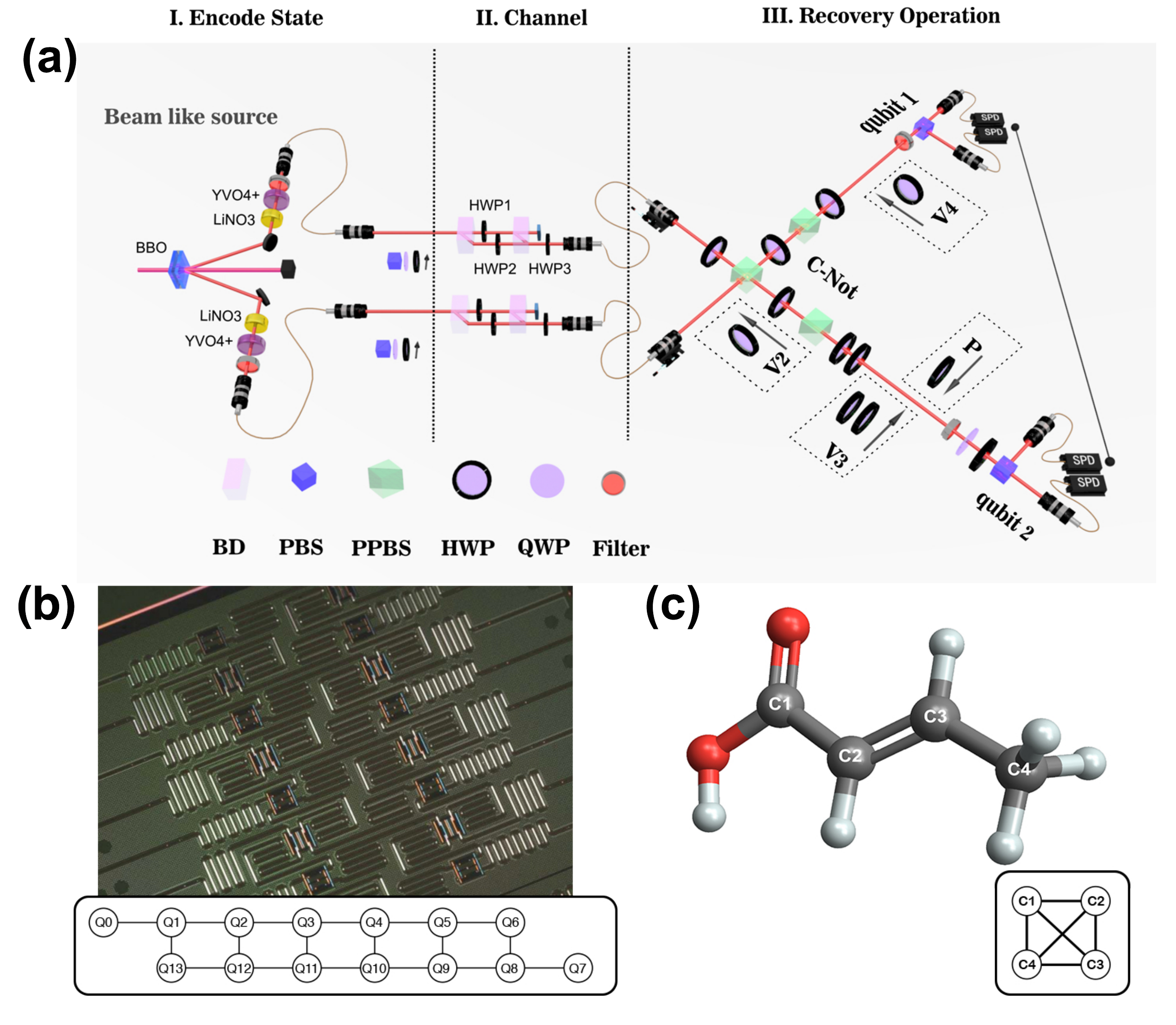}
	\caption{\label{Fig.Device} Illustration of the different quantum
		systems. (a) On the optical platform, we utilize a $390$~nm
		femto-second light to pump a sandwich beamlike phase-matching
		$\beta$-barium-borate (BBO) crystal to generate pairs of
		polarization entangled photon as qubits. (b) IBM Q 16 Melbourne,
		consisting of $14$ superconducting qubits connected via microwave
		resonators, together with the coupling structure. (c) The four
		qubits on the Crotonic acid are given by the spin-$1/2$ nuclear
		spins of ${}^{13}C$.  Each of the four spins couples to the
		other three.}
\end{figure}

As shown in Fig.~\ref{Fig.Device}~(a), a $390$~nm femto-second light
(frequency-doubled from a $780$~nm mode-lock Ti:sapphire pulsed laser
with a pulse width of $150$~fs and repetition rate $76$~MHz) pumps a
sandwich beamlike phase-matching $\beta$-barium-borate (BBO) crystal
to generate pairs of polarization entangled photon
$\frac{1}{\sqrt{2}}(\ket{HV}+\ket{VH})$ in the spontaneous parametric down-conversion (SPDC) process.  
Based on the entangled photons
source, we can prepare the desired encoded state for the six different
states (see the Supplementary Material) by using polarization beam
splitters (PBS), half wave plates (HWP), and quarter wave plates
(QWP). The detailed configurations are given in Table~\ref{optical1T} in the Supplementary Materials.

As illustrated in the middle part of Fig.~\ref{Fig.Device}~(a), for
the optical platform we use an interferometer to implement the
detected amplitude damping channel \cite{kevin2012}.  After passing
through the first beam displacer (BD), the photons with horizontal
polarization ($H$) and vertical polarization ($V$) are parallelly
displaced with respect to each other \cite{work1, work2}. For the
operator $A_0$, the amount of damping $\gamma$ is adjusted by rotating
HWP1 placed between two BDs by the angle $\theta$, with
$\sin^{2}2\theta=1-\gamma$. Meanwhile HWP2 and HWP3 are rotated by
$45^{\circ}$ to perform the bit-flip operator. Regarding $A_1$, HWP1
is rotated by $\theta$, where $\gamma=\sin^{2}2\theta$. Both HWP2 and
HWP3 are set at $0^{\circ}$ to remove the horizontally
polarized photon (part II in Fig.~\ref{Fig.Device}~(a)).  Hence, the
two interferometers together (middle part of
Fig.~\ref{Fig.Device}~(a)) can simulate the four error pattern:
$A_0A_0$, $A_0A_1$, $A_1A_0$, and $A_1A_1$. In the error correction
part, we use the method of Refs.~\citenum{CNOT1,CNOT2,CNOT3} to implement an
all-optical CNOT gate, which is constructed by partially
polarizing beam splitters (PPBS) and HWPs. To quantify the quality of
the CNOT gate, we perform quantum process tomography showing that the
fidelity between the implemented and the ideal gate is about $88.5 \%$
\cite{PTomo}. The errors are mainly caused by the mode mismatch of the
Hong-Ou-Mandel (HOM) interferometer.  In our experiment, the error
patterns and the corresponding recovery operations are given in
Table~\ref{Table.1}, where the gates $H$ and $X$ can be easily
realized by rotating the HWP by $22.5^{\circ}$ and $45^{\circ}$
respectively. The detailed information about the case without error
correction is given in the Supplementary Materials.

\begin{figure}
  \centering
    \includegraphics[width=6.5in]{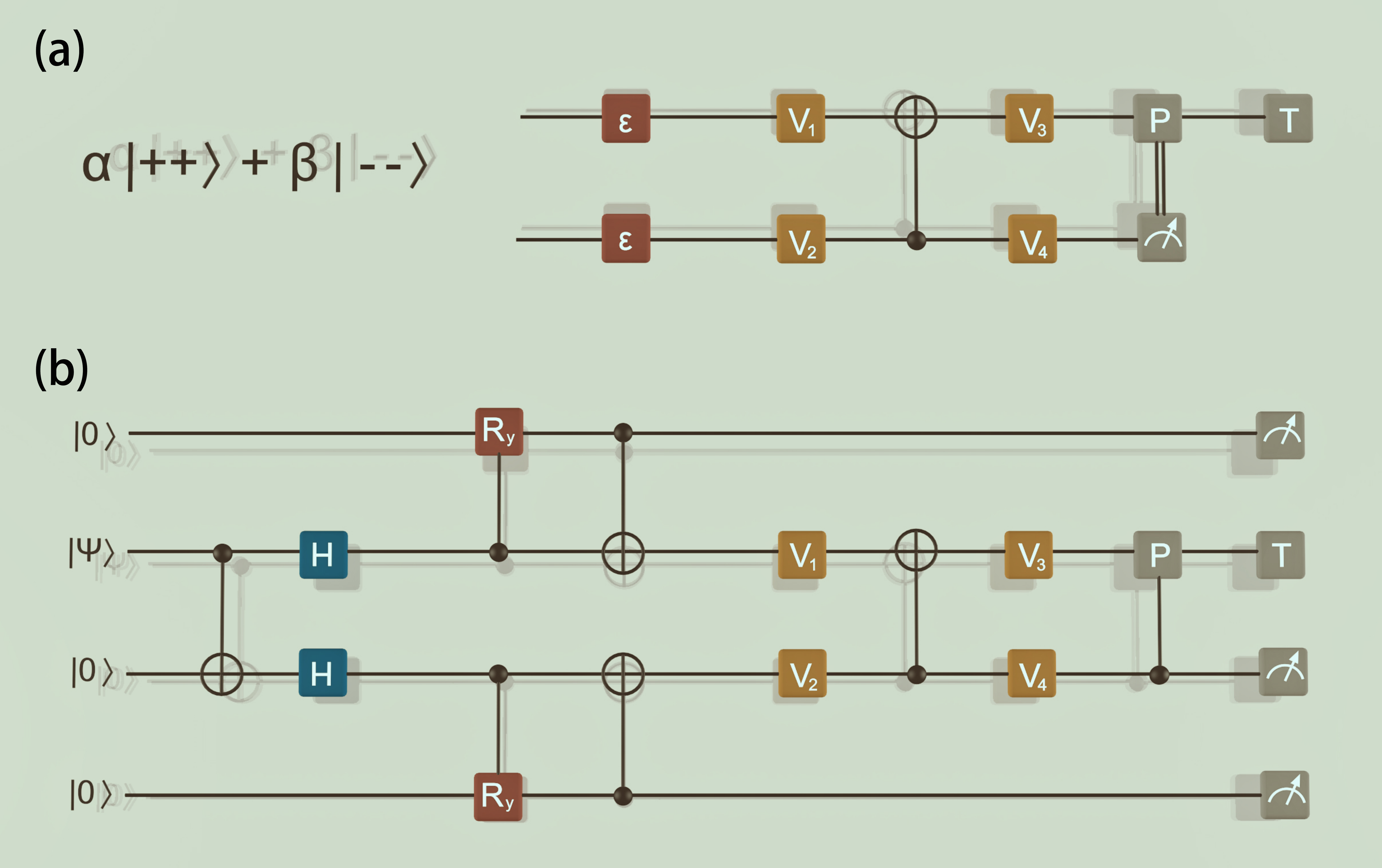}
    \caption{\label{Fig.circuit}Quantum circuits for our
      experiments. (a) The optical platform. After generating pairs of
      entangled photon, we prepare the desired encoded state
      $\alpha|++\rangle+\beta|--\rangle$ with polarization beam splitters
      and wave plates, see Table~\ref{Table.1}. The detected amplitude damping
      channel is depicted by $\varepsilon$. In the error correction
      part, we implement Standard Correction~A (see Supplemental
      Material) using four single qubit rotations and a CNOT based on
      a HOM interferometer.  For the reconstruction of the decoded
      state, we use post-selection on the other qubits. (b) The
      circuit for IBM Q and the NMR system.  Artificial amplitude
      damping channels are implemented by a controlled-$y$-rotation
      from the encoded qubits to the ancillas and the subsequent CNOT
      gates.  Measuring the ancilla qubits at the end reveals which
      error has occurred.  The single-qubit gates $V_1$, $V_2$, $V_3$,
      and $V_4$ in the recovery circuit depend on the particular error.
      To simplify the circuit, we run experiments with all settings
      and use post-selection on the corresponding measurement results
      of the ancillas.  At the end, we use single-qubit state
      tomography on the second qubit to reconstruct the density
      matrix.}
\end{figure}

On IBM Q and the NMR system we use two ancilla qubits to implement the
two-qubit detected amplitude channel.  The qubits of \textit{IBM Q 16
  Melbourne} and the Crotonic acid (see Fig.~\ref{Fig.Device}~(b) and
(c), resp.) meet the required coupling structure (other quantum chips
from IBM Q do not match this connectivity map). To be more concrete,
$Q_5$, $Q_6$, $Q_8$, $Q_9$ on IBM Q 16 Melbourne are selected because
the average error rates of CNOTs between those qubits are lower than
others.  Generally, there are three parts in the quantum circuit,
encoder, amplitude channel, and recovery circuit (containing the
decoder) in the IBM Q and the NMR experiments, as shown in Fig.~\ref{Fig.circuit}~(b).  First, we prepare the initial state
$|\psi\rangle$ by a single qubit rotation of $Q_5$. A CNOT gate and
two Hadamard gates compose the encoder.  With controlled-$y$-rotation
gates $R_y(\theta)$ acting on the ancillas with the encoded qubits as
control and CNOT gates acting on the encoded qubits, we can simulate
the two-qubit detected amplitude channel
\cite{Nielsen:2011:QCQ:1972505}.  The relation between the damping
ratio $\gamma$ and the rotation angle $\theta$ is
$\gamma=\sin^2(\theta/2)$. Measuring the ancilla qubits reveals which
type of error occurred. If the result is $|0\rangle$, $A_0$ has
occurred on the corrsponding encoded qubit, while $A_1$ has occurred
when the result is $|1\rangle$.  Recovery circuits optimized for
\textit{IBM Q 16 Melbourne} and the NMR system are shown in
Fig.~\ref{Fig.circuit}~(b). To extract the quantum density matrix
of the decoded qubit, we use quantum state tomography (QST) and post
selection (see the Supplemental Material), measuring
the output of the same quantum circuit in different bases. For the IBM
Q experiments, we construct the circuit with three-parameter single
qubit rotation gates $U_3(\theta,\lambda,\phi)$ and CNOT gates. For
the NMR experiments, we generate the pulse sequences of the encoder,
two-qubit amplitude damping channel, and recovery circuit using an optimized shape pulse sequence with a total time of $61$~ms.

\subsection{Experimental Results}
\begin{figure}
  \centering
    \includegraphics[width=5.0in]{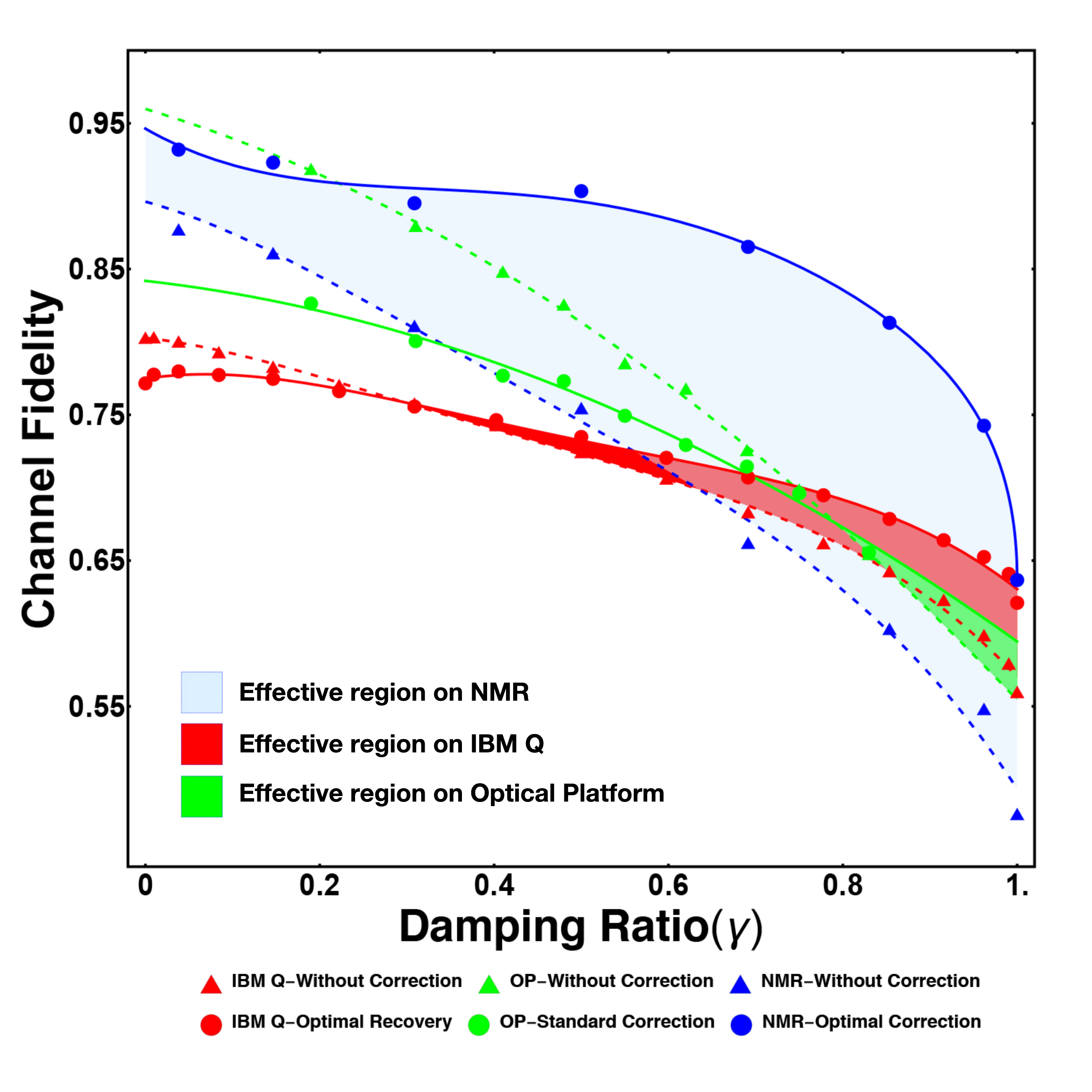}
    \caption{\label{Fig.result} Comparison of the error correction
      capacity on the different systems. The red, green, and blue
      regions charactize the effective region enclosed by the fidelity
      curves for IBM Q, the optical platform (OP), and the NMR system,
      respectively.  The solid and dashed lines are polynomial fits to
      our experiment data. The solid lines represent Optimal Recovery,
      and the dashed lines correspond to the case Without Correction.
      For IBM Q and the optical platform, when the damping ratio
      $\gamma$ is small, Optimal Recovery (resp. Standard Correction~A)
      performs worse initially because of the limited fidelity of the
      additional encoding operations. When the damping ratio $ \gamma
      $ increases, Optimal Recovery reveals its capacity gradually.}
\end{figure}

The main experimental results for the three systems are shown in
Fig.~\ref{Fig.result}.  The fidelity of the effective communication
channel is plotted as a function of the damping parameter
$\gamma$. For the three different systems, we show the effective
regions for which Optimal Recovery (respectively Standard Correction~A
for the optical platform) yields a higher fidelity than using no error
correction.  Without error correction, the optical platform shows a
great advantage in comparison to the other two systems, with the
performance of IBM Q being the lowest. However, with error correction,
the situation changes dramatically.  For the optical platform, the
state fidelity drops already a lot at $\gamma=0$, while adding
error-correcting only slightly reduces the fidelity at $\gamma=0$ for
IBM Q.  Exhibiting the largest effective region (lighter blue), our
error correction scheme exhibits a good performance on the NMR system,
and the maximal improvement at $\gamma\approx 0.6$ reaches
approximately $0.2$. For IBM Q, the improvement (red region) is
smaller, but it is still given for a large range of damping
parameters $\gamma$.  For the optical platform, error correction
improves the overall fidelity only a little for $\gamma> 0.83$

On the optical platform (see Fig.~\ref{figS1} in the Supplementary
Material) we perform experiments with Standard Correction~A and
without correction.  At the mercy of the bad fidelity of implementing
the CNOT by HOM interference, we find that at lower damping
probabilities $(\gamma=0.17 \sim 0.83)$, the fidelity for the state
without correction is larger than with standard correction. However,
if the damping probability $\gamma$ is larger than $0.83$, standard
error correction will be better.  This demonstrates some limited
improvement using quantum error correction. 

Fig.~\ref{figS2} in the Supplementary Material shows the result for
IBM Q averaging $4096$ runs for $16$ sample points.  For
$\gamma\in[0.0, 0.36)$, no correction yields a higher fidelity than
  Optimal Recovery since ``Without Correction'' involves only two
  qubits.  Generally, it is ubiquitous to QEC that the encoded states
  get worse initially as the encoding operations reduce the
  fidelity. The blue star plotted at $\gamma=0.36$ in
  Fig.~\ref{figS2}~(a) indicates the intersection when the overall
  fidelity of ``Optimal Recovery'' equals ``Without Correction''.
  When the damping parameter $\gamma$ increases, both Standard
  Correction~A and Standard Correction~B show the capacity of error
  correction, but neither outperforms Optimal Recovery.

The results for the NMR system are show in Fig.~\ref{figS3}. Optimal
Recovery, Standard Correction~A, as well as Standard Correction~B show
substantial improvements in comparison to Without Correction,
indicating the power of quantum error correction. Furthermore, the
state fidelity curves for Standard Correction~A and Standard
Correction~B exhibit faster decay than the curve for Optimal Recovery,
revealing that Optimal Recovery is indeed the best error correction
scheme for the detected amplitude damping channel, which matches the
theoretical result.

\section{Discussion}
The experiments mainly demonstrate the potential to realize quantum
error correction on a quantum computer in the NISQ era by implementing
an optimal error-correcting code for detected amplitude damping on IBM
Q, an optical platform, and an NMR system.  All experiments provide
evidence that the advantage of quantum error correction can even be
revealed on a present quantum computer, with only a few qubits and
faulty quantum gates. For all three systems, Optimal Recovery shows
eventually an improvement in comparison to Without Correction.  On the
other hand, for small damping parameters $\gamma$, the correction
scheme does not come into effect on the optical platform and on IBM Q.

Our experiments also reveal the underlying relation between the
ability of quantum control and the performance of quantum error
correction.  In a typical quantum information process, quantum errors
mainly stem from imprecise readout, decoherence, and faulty CNOT
gates.

Firstly, we consider the influence of imprecise readout.  The readout
error for the optical platform and the NMR system can be neglected
because for both the precision of readout is close to $99.9~\%$.  On
the IBM Q platform, however, the average error rate of readout is
nearly $5.0~\%$, see Table~\ref{Table.2} in the Supplementary
Material. Apparently, the readout error only contributes a fraction of
the entire infidelity in our experiments.

The qubit quality, especially the coherence time, is also an important
factor for the performance of the quantum error-correcting code.  A rough
estimate for the state $\frac{1}{\sqrt{2}}(|0\rangle+|1\rangle)$ shows
that the decoherence error contribution from $T_2$ for IBM Q and the
NMR system are $3.5~\%$ and $9.9~\%$, respectively (see the
Supplementary Material). Therefore, the decoherence is not the main
source of the infidelity in IBM Q experiments, but may cause the dominant
error in the NMR system.

For the optical system, substantial infidelity is contributed by the
CNOT based on HOM interference. When adding the CNOT to the recovery
circuit, the total shot numbers of photons will be suppressed by the
PPBS crystal. We denote the phenomenon by shot loss. If we use ideal
probabilities that the errors happen instead of the real probabilties
in the experiments to reconstruct the effective density matrix, the
correcting effect will enhance, see Fig.~\ref{figS4}. A similar effect
occurs for IBM Q because of cross-resonance CNOT gates. This
phenomenon stems from $ZZ$-crosstalk in the superconducting qubit
chips \cite{PhysRevLett.122.080504}. To reconstruct the density matrix
with ideal probabilties, even Standard Correction A shows the
capcity to improve the channel fidelity in Fig.~\ref{figS4}. However,
in the NMR experiments, we use the GRAPE algorithm to generate the
total pulse sequence with a precision of $99.9~\%$, which gives a
great improvement to CNOTs and other operations.

In conclusion, our experiments demonstrate that the quality of
CNOT mainly influences the performance of quantum error correction. CNOT
operations, at the core of both encoder and decoder, play a unique role to
generate entanglement in both quantum error correction and quantum
computing.  Our results motivate further investigations to improve
the precision of CNOT operations and indicate the route towards viable
quantum error correction in the NISQ era.

\begin{methods}
\subsection{Standard Correction A/B}
For the two-qubit code given by Eq.~\eqref{Eq.2}, Standard Correction A/B protocols can be derived as follows:
\begin{itemize}
	\item If $A_0 A_1$ (or $A_1 A_0$) happens, discard the qubit on
	which $A_1$ happened.  On the other qubit, apply $X$ to compensate
	for the phase error introduced by $A_1$ acting differently on
	$|+\rangle$ and $|-\rangle$.
	\item If $A_0 A_0$ happens, directly decode the two qubits.
	\item If $ A_1 A_1 $ happens, the quantum state $|\psi\rangle$ is
	converted to the state $ |00\rangle $.  To maximize the fidelity,
	we transform it to an equally weighted superposition state
	$\frac{1}{\sqrt{2}}(|0\rangle+|1\rangle)$. There are two different
	schemes to create an equally weighted superposition state
        which we refer to as Standard Correction A/B (see Table.~\ref{Table.1}), respectively.
\end{itemize}

\subsection{Optimal Recovery}
The Optimal Recovery operators are derived in the Supplementary
Material.  We find a pair of recovery operations $V_3$ and $V_4$ that
can be implemented by Pauli gates, the Hadamard gate, a CNOT gate and
general single-qubit three-parameter rotation gates.  The two recovery
operations have the form $ V_3=U^\dagger_1 H $ and $ V_4=HU^\dagger_2 X $, where
\begin{equation}
U_1=\dfrac{1}{\sqrt{(1+t)^2+(1-s)^2}} \begin{pmatrix}-t-1 & s-1\\ -s+1 & -t-1 \end{pmatrix} \ ,
\end{equation}
\begin{equation}
U_2=\dfrac{1}{\sqrt{(1+t)^2+(1-s)^2}} \begin{pmatrix}-s+1 & t+1\\ -t+1 & -s+1 \end{pmatrix} \ ,
\end{equation}
where the parameters $s$ and $t$ are given by
\begin{equation}
s=\dfrac{\sqrt{2}}{\sqrt{1+(1-\gamma)^2}}\quad\text{and}\quad t=\dfrac{\sqrt{2}(1-\gamma)}{\sqrt{1+(1-\gamma)^2}}\ .
\end{equation}

The general setup of the circuit for both standard correction and
optimal recovery is shown in Fig.~\ref{Fig.Model}. Information on the
specific circuits is given in Table~\ref{Table.1} and
Fig.~\ref{Fig.circuit}.
\end{methods}



\bibliography{sample}


\begin{addendum}
 \item Q.H. Guo and Y.-Y. Zhao contributed equally to this work. We thank Dawei Lu, Shuming Cheng, Kevin Resch, Runyao Duan for fruitful discussions. Y.-Y. Zhao is supported by the National Natural Science Foundation for the Youth of China (Grants No.11804410). M. Grassl acknowledges partial support by the Foundation for Polish Science (IRAP project, ICTQT, contract no. 2018/MAB/5, co-financed by EU within the Smart Growth Operational Programme). G.-Y. Xiang is supported by the National Natural Science Foundation of China (Grants No.11574291, 11774334). T. Xin is supported by National Natural Science Foundation of China (Grants No. 11905099 and No. U1801661), and Guangdong Basic and Applied Basic Research Foundation (Grant No. 2019A1515011383). Z.-Q. Yin is supported by National Natural Science Foundation of China (Grants No. 61771278) and Beijing Institute of Technology Research Fund Program for Young Scholars.
 
 We gratefully acknowledge use of the IBM Q for this work. The views expressed are those of the authors and do not reflect the official policy or position of IBM or the IBM Q team.
 \item[Competing Interests] The authors declare that they have no
competing financial interests.
 \item[Correspondence] Correspondence and requests for materials should be addressed to G.-Y.X. (email: gyxiang@ustc.edu.cn), T.X. (email: xint@sustech.edu.cn) or B.Z. (email: zengb@ust.hk).

\end{addendum}
\pagebreak
\widetext
\begin{center}
	\textbf{\large Supplemental Material:\\ Testing a Quantum Error-Correcting Code on Various Platforms}
\end{center}
\setcounter{section}{0}
\setcounter{equation}{0}
\setcounter{figure}{0}
\setcounter{table}{0}
\setcounter{page}{1}
\makeatletter
\renewcommand{\theequation}{S\arabic{equation}}
\renewcommand{\thefigure}{S\arabic{figure}}
\renewcommand{\bibnumfmt}[1]{[S#1]}
\renewcommand{\citenumfont}[1]{S#1}

\section{The Amplitude Damping Channel}
The amplitude damping channel is an important model that describes
spontaneous emission and the loss of energy in quantum communication
\cite{Nielsen:2011:QCQ:1972505,PhysRevLett.86.4402}. Generally, the
effect of s channel on a quantum state is represented by a completely
positive, trace-preserving linear map acting on a density matrix
$\rho$. Such a map can be expressed in the Kraus representation
$\mathcal{A}(\rho)=\sum_i A_i \rho A_i^\dagger$ with $\sum_i
A_i^\dagger A_i=I$. The Kraus representation of the single-qubit
amplitude damping channel has the form
\begin{equation}\label{eq1}
\rho \to \rho'=\mathcal{A}_{AD}(\rho)=A_0 \rho A_0^\dagger+A_1 \rho A_1^\dagger \ ,
\end{equation}
where the Kraus operators are
\begin{equation}\label{eq:ADchannel}
A_0=\begin{pmatrix} 1   & 0 \\ 0 & \sqrt{1-\gamma} \end{pmatrix} \ \text{and} \ A_1=\begin{pmatrix} 0   & \sqrt{\gamma} \\ 0 & 0 \end{pmatrix} \ .
\end{equation}
The operation $A_1$ maps the state $|1\rangle$ to the state
$|0\rangle$, corresponding to the entire loss of energy of a qubit,
e.g., due the the spontaneous emission of a photon. The operation
$A_0$ does not change the state $|0\rangle$, but reduces the amplitude
of the state $|1\rangle$.  For the detected amplitude damping channel,
the additional classical information is available which of the two
cases has occurred.  To simulate a detected amplitude damping channel,
we have to apply the operation $A_0$ or $A_1$ to the state, which are
are non-unitary local operation.  hence we need
local filters \cite{verstraete2001local} in quantum optics or add
ancillas for the superconducting circuit and NMR
system \cite{Nielsen:2011:QCQ:1972505}, making use of post selection.

\section{Post Selection and State Tomography}
For each qubit, there are the two possible errors $A_0$ and $A_1$,
which in our simulation of the channel can be distinguished by the
outcome of measuring the ancillas.  In total, four different error
patterns $A_iA_j=A_i\otimes A_j$ may occur on the encoded bipartite
state.  To be more concrete, assume that the initial state
$|\psi\rangle$ was prepared on qubit $A$. Then the encoding operation
gives the encoded two-qubit state $\rho_{A,B}$ on qubits $A$ and $B$.
Suppose that the channels act on both qubits of the encoded state
$\rho_{A,B}$.  Then we will obtain the density matrix $\rho_{ij}'$
with probability $p_{ij}$, where
\begin{equation}
\rho_{ij} '= \dfrac{A_i A_j \rho_{A,B} A_i^\dagger A_j^\dagger}{\mathrm{Tr}(A_i A_j \rho_{A,B} A_i^\dagger A_j^\dagger)}  \ ,
\end{equation}
\begin{equation}
p_{ij}=\mathrm{Tr}(A_i A_j \rho_{A,B}A_i^\dagger A_j^\dagger) \ ,
\end{equation}
for $ i,j=0,1 $.  To recover the information of the quantum state,
we decode the density matrix $\rho_{ij}'$ and then trace out  qubit $B$.
The final reduced density matrix $\rho_A'$ can be obtained by
taking the weighted sum of the results of single qubit state tomography on
the qubit $A$ for the four cases $i,j=0,1$. Therefore,  the final density matrix $\rho_A'$ has the form
\begin{equation}\label{Eq.3}
\rho_A' = \sum_{i,j}p_{ij}\mathrm{Tr}_{B}  \bigg[\mathcal{D}\bigg(\dfrac{A_i A_j \rho_{A,B}A_i^\dagger A_j^\dagger}{\mathrm{Tr}(A_i A_j \rho_{A,B}A_i^\dagger A_j^\dagger)}\bigg) \bigg] \ ,
\end{equation}
where $\mathcal{D}$ denotes the decoding operation. To measure
the distance between the final density matrix $ \rho_A' $ and the
initial quantum state $ \rho_A=|\psi\rangle\langle \psi| $, we use the
fidelity $ F(\rho_A,\rho_A') $, defined as
\begin{equation}
F(\rho, \rho_A' )=\text{Tr} (\rho_A^{1/2} \rho_A' \rho_A^{1/2})=\langle\psi|\rho_A'|\psi\rangle.
\end{equation}

In order to compute the average fidelity of the whole communication
system, it suffices to consider the six input states
\begin{alignat}{5}
&\ket{H}=\ket{0}, &\quad& \ket{V}=\ket{1}\\
&\ket{D}=\frac{1}{\sqrt{2}}(\ket{0}+\ket{1}), &&\ket{A}=\frac{1}{\sqrt{2}}(\ket{0}-\ket{1})\\
&\ket{R}=\frac{1}{\sqrt{2}}(\ket{0}+i\ket{1}), &&\ket{L}=\frac{1}{\sqrt{2}}(\ket{0}-i\ket{1})
\end{alignat}
which are the states of three mutually unbiased bases forming a $2$-design.

\section{Optimal Recovery}
We encode the state $|0\rangle$ as $|++\rangle$ and the state
$|1\rangle$ as$|--\rangle$. Then the encoding isometry is given
by $\rho\mapsto E\rho E^\dagger$, where
\begin{equation}
E=|++\rangle \langle 0|+|--\rangle\langle 1|
\end{equation}
Combining encoding and the amplitude damping channel, we obtain Kraus operators
\begin{equation}
t_{ij}=(A_i \otimes A_j)E,
\end{equation}
where $i,j=0,1$.

Writing $t_{ij}$ in its polar decomposition
\begin{equation}\label{S5}
t_{ij}=v_{ij}|t_{ij}|,
\end{equation}
where $|t_{ij}|=\sqrt{t_{ij}^\dagger t_{ij}}$  and $v_{ij}$ are
isometries, the recovery operations are given by
\begin{equation}
R_{ij}(\rho)=v^\dagger_{ij}\rho v_{ij}+\rho_{ij} \text{tr}(\rho(I-v_{ij}v_{ij}^\dagger)),
\end{equation}
where $\rho_{ij}$ are arbitrary single qubit states.

As $\rho_{ij}\text{tr}(\rho(I-v_{ij}v_{ij}^\dagger))=0 $ for all
$\rho=t_{ij}\rho_s t_{ij}^\dagger$ (where $\rho_s$ is an arbitrary
single qubit state), we will only need to implement $v_{ij}^\dagger$,
together with an arbitrary completion to a POVM. The isometries $v_{ij}$ are given by
\begin{equation}
v_{00}=\frac{1}{2}\begin{pmatrix}
\frac{\sqrt{2}}{\sqrt{1+(1-\gamma)^2}} & \frac{\sqrt{2}}{\sqrt{1+(1-\gamma)^2}}\\
1 & -1 \\
1 & -1 \\
\frac{\sqrt{2}(1-\gamma)}{\sqrt{1+(1-\gamma)^2}} & \frac{\sqrt{2}(1-\gamma)}{\sqrt{1+(1-\gamma)^2}}
\end{pmatrix},
v_{01}=\frac{1}{\sqrt{2}}\begin{pmatrix}
1 & -1\\
0 & 0 \\
1 & 1 \\
0 & 0
\end{pmatrix}
\end{equation}

\begin{equation}
v_{10}=\frac{1}{\sqrt{2}}\begin{pmatrix}
1 & -1\\
1 & 1 \\
0 & 0 \\
0 & 0
\end{pmatrix},
v_{11}=\frac{1}{\sqrt{2}}\begin{pmatrix}
1 & 1\\
1 & -1 \\
0 & 0 \\
0 & 0
\end{pmatrix}
\end{equation}
It can be shown that $v_{00}^\dagger$ cannot be directly implemented
in a unitary way, i.e., by a two-qubit unitary followed by tracing
out one qubit, using a single CNOT operation.

We finally obtain the single-qubit gates $V_i$ for the error
correction circuit shown in Fig.~\ref{Fig.circuit}, 
\begin{equation}
V_1=\dfrac{1}{\sqrt{2}} \begin{pmatrix}1 & -1\\ 1 & 1 \end{pmatrix} \ , V_2=\dfrac{1}{\sqrt{2}} \begin{pmatrix}1 & 1\\ -1 & 1 \end{pmatrix} \ ,
\end{equation}
\begin{equation}\label{key}
V_3= U^\dagger_1 H\,, V_4=HU^\dagger_2 X\,,
\end{equation}
where
\begin{equation}
U_1=\dfrac{1}{\sqrt{(1+t)^2+(1-s)^2}} \begin{pmatrix}-t-1 & s-1\\ -s+1 & -t-1 \end{pmatrix} \ ,
\end{equation}
\begin{equation}
U_2=\dfrac{1}{\sqrt{(1+t)^2+(1-s)^2}} \begin{pmatrix}-s+1 & t+1\\ -t+1 & -s+1 \end{pmatrix} \ ,
\end{equation}
where the parameters $s$ and $t$ are given by
\begin{equation}
s=\dfrac{\sqrt{2}}{\sqrt{1+(1-\gamma)^2}}, t=\dfrac{\sqrt{2}(1-\gamma)}{\sqrt{1+(1-\gamma)^2}}\ .
\end{equation}

In order to implement the decoding isometry
\begin{equation}
D=E^\dagger=|0\rangle\langle ++|+|1\rangle \langle --|,
\end{equation}
one can measure the second qubit in the computational basis, followed
by a Hadamard transformation $H$ on the first qubit. If $|0\rangle$ is
obtained in the measurement, then the state of the first qubit is
$\alpha'|0\rangle+\beta'|1\rangle$.  If $|1\rangle$ is obtained, we
have $ \alpha'|0\rangle-\beta'|1\rangle$, so a $Z$-gate has to be
applied before the single-qubit state tomography.

\section{Preparation of the Encoded States for the Optical Platform}
In Table~\ref{optical1T}, we list the optical elements needed to
prepare the six different encoded states from a polarization entangled
two-phton state.
\begin{table}
	\centering
	\begin{tabular}{|c|c|c|}
		\hline
		\textbf{Initial state} & \textbf{Optical elements (qubit 1)}& \textbf{Optical elements (qubit 2)} \\
		\hline
		$\frac{1}{\sqrt{2}}(\ket{0}+\ket{1})$  & HWP@$45^{\circ}$ &None \\
		\hline
		$\frac{1}{\sqrt{2}}(\ket{0}-\ket{1})$  & None & None \\
		\hline
		$\ket{0}$  & PBS, HWP@$22.5^{\circ}$ & PBS, HWP@$22.5^{\circ}$ \\
		\hline
		$\ket{1}$  & PBS, HWP@$-22.5^{\circ}$ & PBS, HWP@$-22.5^{\circ}$\\
		\hline
		$\frac{1}{\sqrt{2}}(\ket{0}+i\ket{1})$ & QWP@$0^{\circ}$, HWP@$22.5^{\circ}$ & HWP@$22.5^{\circ}$ \\
		\hline
		$\frac{1}{\sqrt{2}}(\ket{0}-i\ket{1})$ & QWP@$90^{\circ}$, HWP@$22.5^{\circ}$ & HWP@$22.5^{\circ}$ \\
		\hline
	\end{tabular}
	\caption{Optical elements needed to prepare the various
          encoded states from a polarization-entangled two-photon state.}\label{optical1T}
\end{table}

\section{Recovery Operations}
In Table~\ref{Table.1} we list the local operations $V_i$ for the
recovery operation (see Fig.~\ref{Fig.circuit}). For Optimal Recovery,
only the operations for the case $A_0 A_0$ are different.
\begin{table}
  \centering
  \begin{tabular}{|l|l|l|l|l|l|l|}
    \hline
    \textbf{Correction Type}     & \textbf{Error Pattern} & $V_1$ & $V_2$ & $V_3$ & $V_4$ & $P$ \\ \hline
     Standard Correction& $A_0 A_1$ & $I$ & $I$ & $HX$          & $I$ & $I$ \\ \hline
                           & $A_1 A_0$ & $I$ & $I$ & $HX$          & $H$ & $X$ \\ \hline
                           \ \ \ \ \ \ \ \ \ \ \ \ \ (A)& $A_1 A_1$ & $I$ & $H$ & $I$           & $I$ & $I$ \\ \hline
                           \ \ \ \ \ \ \ \ \ \ \ \ \ (B)& $A_1 A_1$ & $I$ & $I$ & $H$           & $I$ & $I$ \\ \hline
                           & $A_0 A_0$ & $I$ & $I$ & $H$           & $H$ & $I$ \\ \hline
       Optimal Recovery    & $A_0 A_0$ & $I$ & $H$ & $U^\dagger_1 H$ &$  HU^\dagger_2 X$    & $Z$\\ \hline
	\end{tabular}
	\caption{\label{Table.1} Recovery operations for the general
          setup. The table lists the specific setup for the correction and
          decoding circuit for the corresponding error patterns.}
	\label{my-label}
\end{table}
\clearpage

\section{Additional Information on the optical platform, IBM Q, and the NMR System}

\subsection{Detected-jump channel without correction}
\begin{figure}
	\centering
	\includegraphics[width=6.0in]{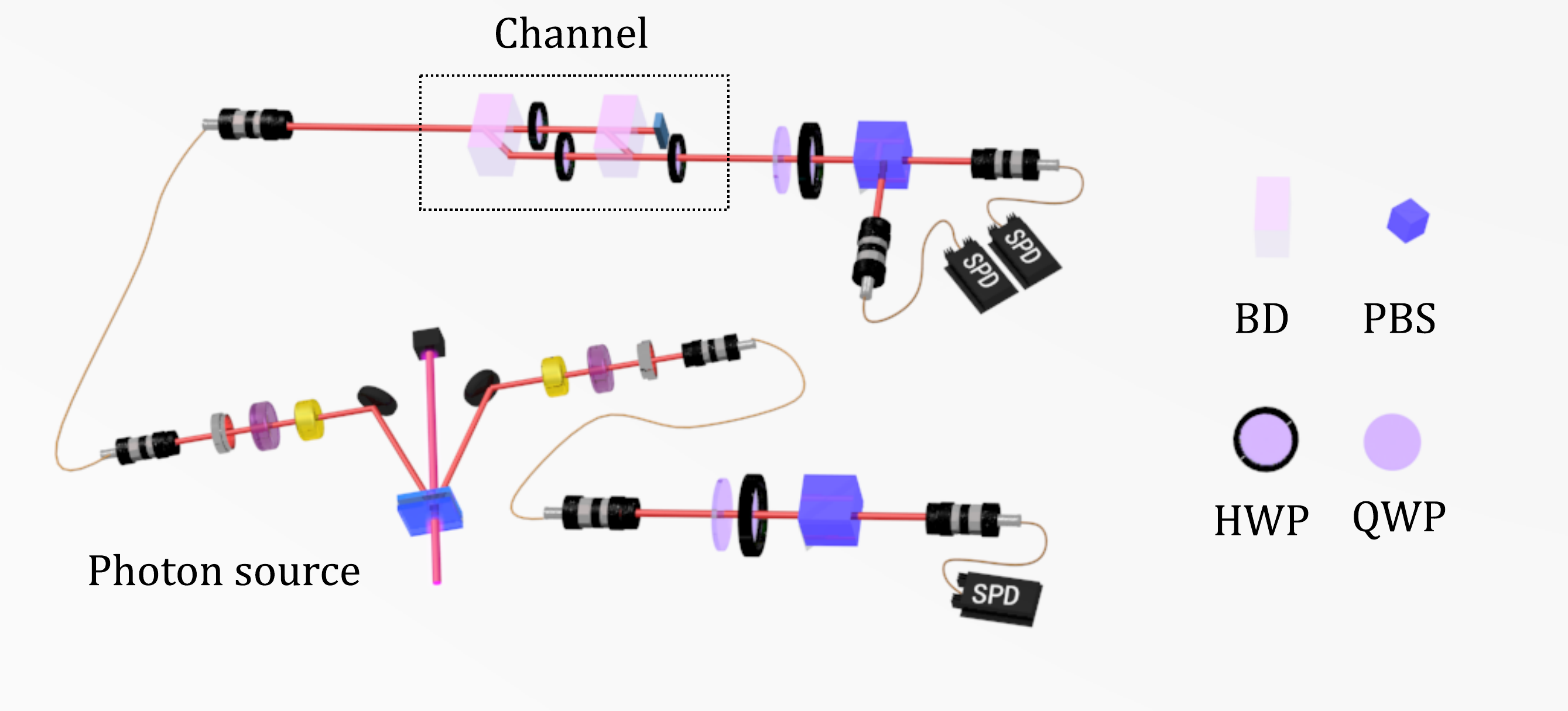}
	\caption{\label{setupWoC} Experimental setup for the process without correction (optical platform)}
\end{figure}

The experimental setup for the case without correction is given in
Fig.~\ref{setupWoC}. The entangled photon state
$1/\sqrt{2}(|HH\rangle+|VV\rangle)$ is prepared through spontaneous
parametric down-conversion (SPDC)
which is similar to the photon source in Fig.~\ref{Fig.Device}~(a) in
the main text. Then one photon of each photon pair is sent to the
detected amplitude channel with the other one providing a trigger
signal. Here we test the six single qubit states $|0\rangle$,
$|1\rangle$, $1/\sqrt{2}(|0\rangle+|1\rangle)$,
$1/\sqrt{2}(|0\rangle-|1\rangle)$, $1/\sqrt{2}(|0\rangle+i|1\rangle)$
and $1/\sqrt{2}(|0\rangle-i|1\rangle)$, which can be prepared by
projecting the trigger photon onto $|H\rangle$, $|V\rangle$,
$1/\sqrt{2}(|H\rangle+|V\rangle)$, $1/\sqrt{2}(|H\rangle-|V\rangle)$,
$1/\sqrt{2}(|H\rangle+i|V\rangle)$, and
$1/\sqrt{2}(|H\rangle-i|V\rangle)$, respectively. At last, standard
single qubit tomography is performed on the photon passing through the
channel.

\subsection{Error probabilities for IBM Q}
In Table~\ref{Table.2}, we list the error probabilities for IBM Q.
\begin{table}
	\centering
	\begin{tabular}{|l|l|l|l|l|}
		\hline
		\textbf{Error Type}                & $Q_5$  & $Q_6$  & $Q_8$ &  $Q_9$   \\ \hline
		single-qubit gate error ($10^{-3}$) & 2.36   & 1.73   & 1.81   & 3.44   \\ \hline
		readout error ($10^{-2}$)           & 4.30   & 3.88   & 3.14   & 7.72   \\ \hline
   		                                   & CX5\_6 & CX6\_8 & CX5\_9 & CX9\_8 \\ \hline
		multi-qubit gate error ($10^{-2}$)  & 6.90   & 5.63   & 6.49   & 5.36   \\ \hline
	\end{tabular}
	\caption{\label{Table.2} Error probabilities for \textit{IBM Q 16 Melbourne}: gate-error and
          and readout-error information for qubits $Q_5$, $Q_6$, $Q_8$,
          $Q_9$, as well as the error rates for the control gates CX5\_6, CX5\_9,
          CX6\_8 and CX9\_8, as provided on 2018-12.}
\end{table}

\subsection{Information about NMR system}

In Table~\ref{tab:Crotonic}, we list the frequencies (Hz) and coupling
constants for the Crotonic acid used in the NMR experiment.
\begin{table}
	\centering
	\begin{tabular}{|l|l|l|l|l|}
		\hline
		       & $C_1$       & $C_2$       & $C_3$       & $C_4$       \\ \hline
		$C_1$  & $2560.603$ &             &             &             \\ \hline
		$C_2$  & $41.65$ & $21837.66$ &             &             \\ \hline
		$C_3$  & $1.47$     & $69.73$    & $18494.94$ &             \\ \hline
		$C_4$  & $7.03$ & $1.17$   & $72.35$  & $25144.73$ \\ \hline
	\end{tabular}
	\caption{ Chemical shifts~(Hz) of the $i$th spin and the $J$-coupling
          strength between spins $i$ and $j$ of the Crotonic acid
          used in the NMR experiment.}
	\label{tab:Crotonic}
\end{table}
\clearpage

\section{Estimate of Decoherence Errors}
For a two-qubit gate, depolarization due to thermal noise can be
estimated from the relaxation times $T_1$ and $T_2$ for each
qubit. For the state $|++\rangle$, a simple model shows that the
density matrix evolves as
\begin{equation}
\rho(t)=\frac{1}{4}\begin{pmatrix}
   1 & 0 & 0& e^{-2t/T_2} \\
   0 & 1 & e^{-2t/T_2} &0 \\
   0 & e^{-2t/T_2} & 1 &0 \\
   e^{-2t/T_2} &0 &0& 1
  \end{pmatrix}.
\end{equation}
The infidelity of the first qubit is
\begin{equation}
P_{\text{sys}}(t,T_2) =\frac{1}{2}-\frac{1}{2}e^{-2t/T_2}.
\end{equation}
For \textit{IBM Q 16 Melbourne}, the average time for a CNOT gate is
approximately $348$~ns, while the time for a single qubit rotation is
about $100$~ns. A buffering time of $20$~ns has to be added before and
after each gate. So the total time on IBM Q is nearly $2680$~ns. The
decoherence error estimate for IBM Q is about $3.5~\%$. For the NMR
system, we use GRAPE to generate a pulse with duration time $61$~ms,
with an error estimate of $9.9~\%$.

\newpage
\section{Detailed Results}
\begin{figure}[H]
	\begin{minipage}[t]{1\hsize}
		\centering \vskip-1cm
		\includegraphics[width=4.3in]{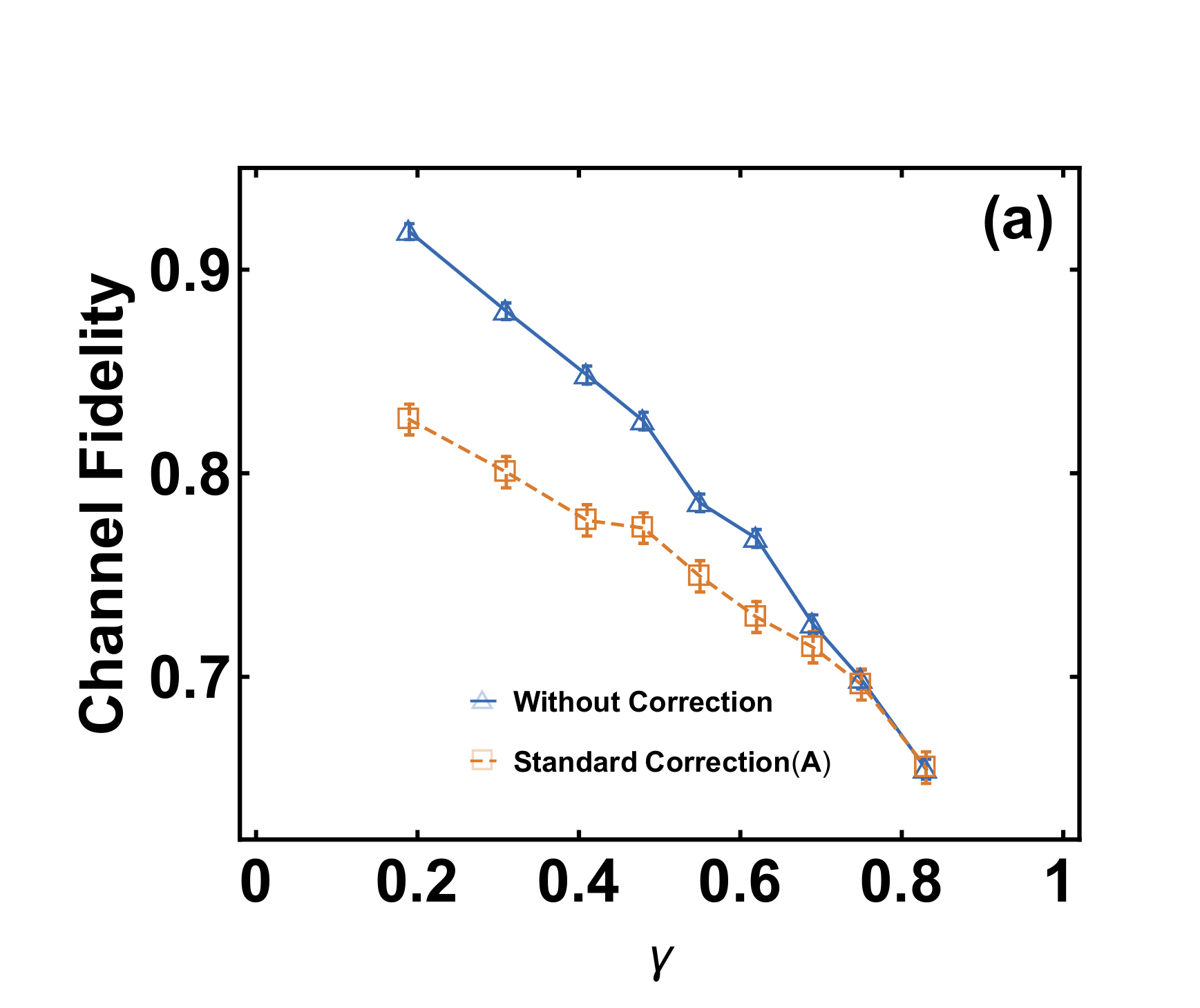}\vskip-5mm
		\caption*{(a) Channel fidelities}
		\label{figS1:side:a}
	\end{minipage}
	\begin{minipage}[t]{0.5\hsize}
		\centering \vskip-1cm
		\includegraphics[width=3.8in]{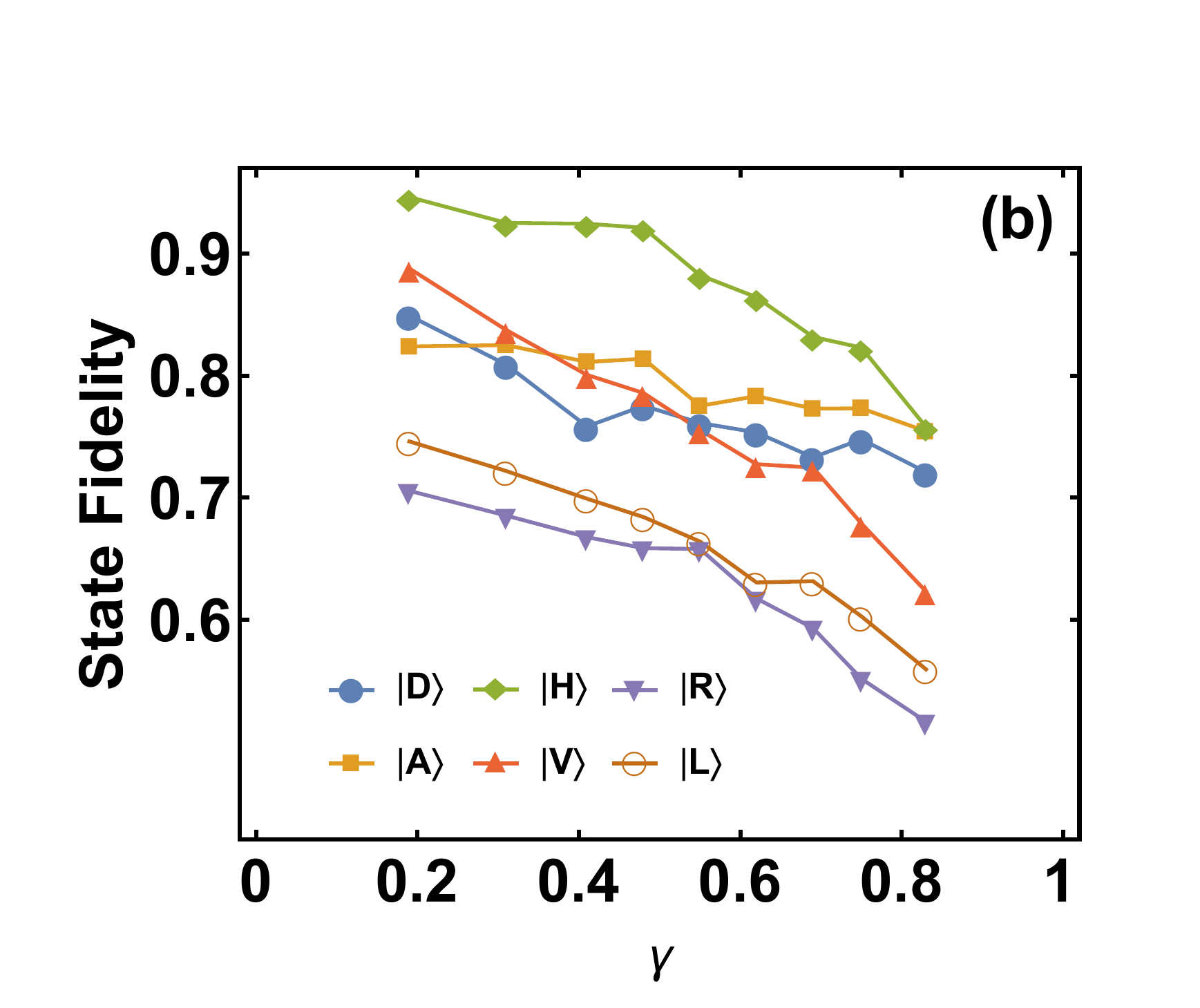}\vskip-5mm
		\caption*{(b) Standard Correction}
		\label{figS1:side:b}
	\end{minipage}
	\begin{minipage}[t]{0.5\hsize}
		\centering \vskip-1cm
		\includegraphics[width=3.8in]{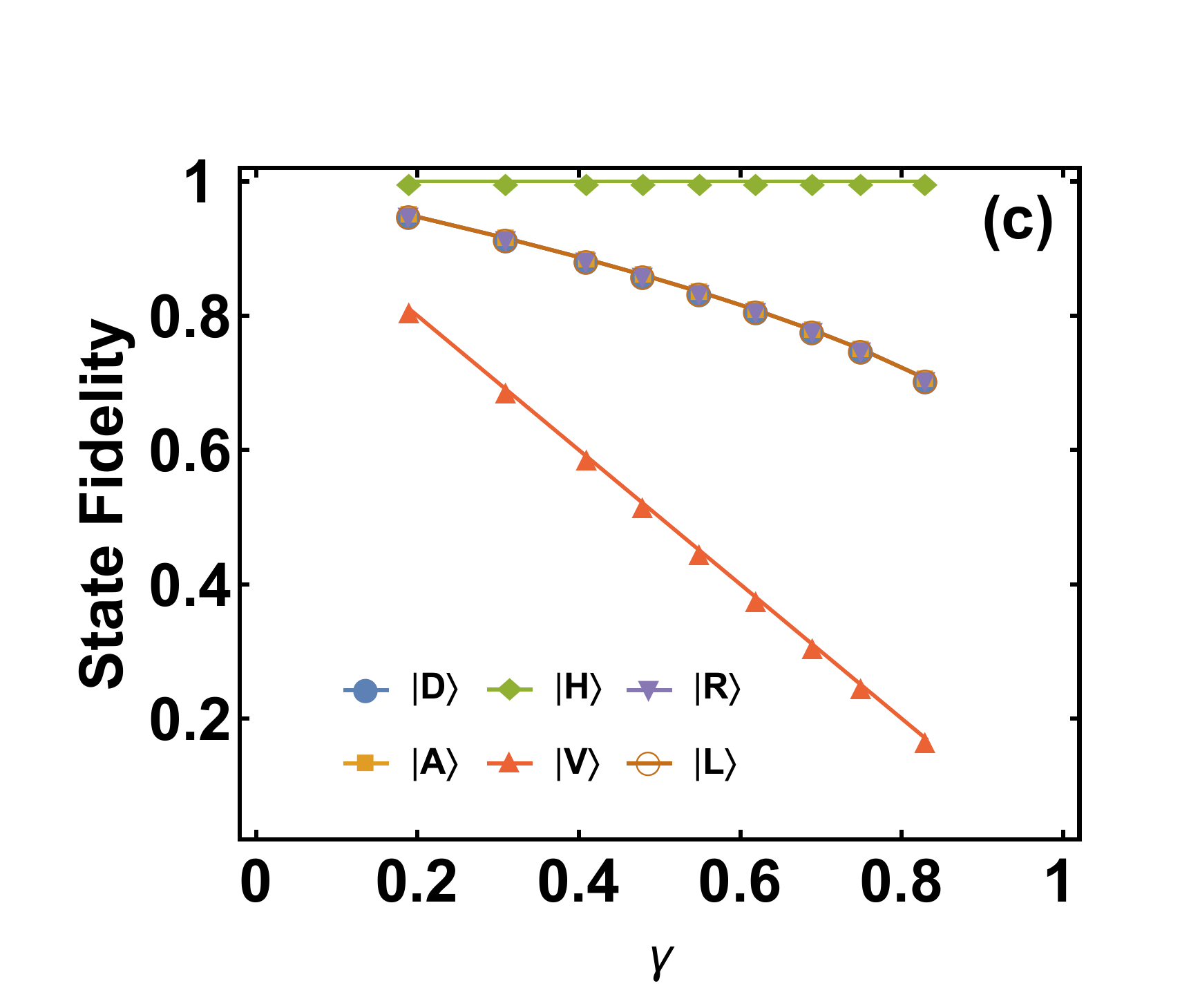}\vskip-5mm
		\caption*{(c) Without Correction}
		\label{figS1:side:c}
	\end{minipage}
	\caption{\label{figS1}Experimental results from the optical
          platform. Standard Correction takes effect until $\gamma$
          reaches $0.83$. Error bars are obtained by Monte Carlo
          simulation (10000 shots).}
\end{figure}
\begin{figure}[H]
	\begin{minipage}[t]{1\linewidth}
		\centering \vskip-1cm
		\includegraphics[width=4.3in]{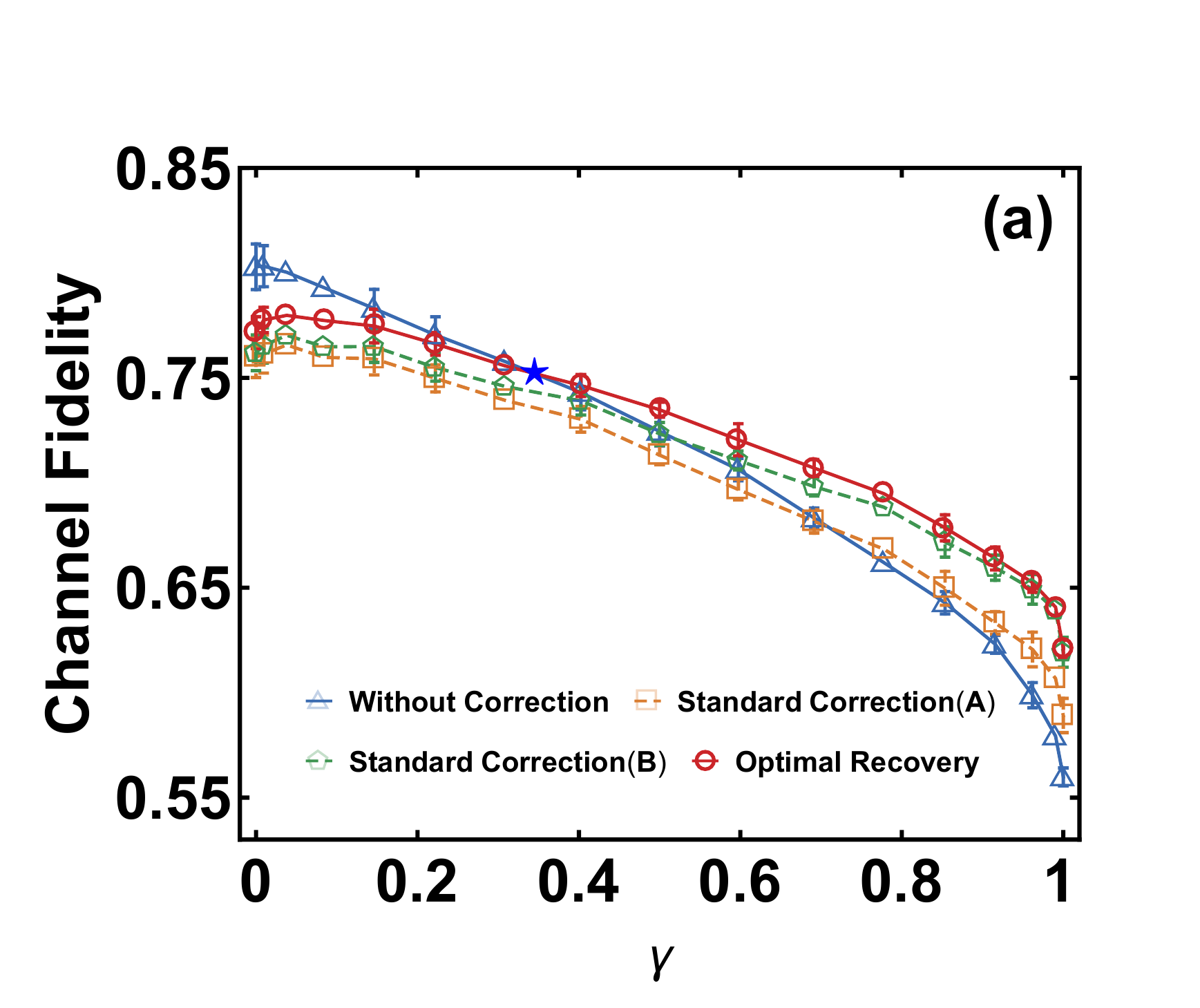}\vskip-5mm
		\caption*{(a) Channel fidelities}
		\label{figS2:side:a}
	\end{minipage}
	\begin{minipage}[t]{0.52\linewidth}
		\centering \vskip-1cm
		\includegraphics[width=3.8in]{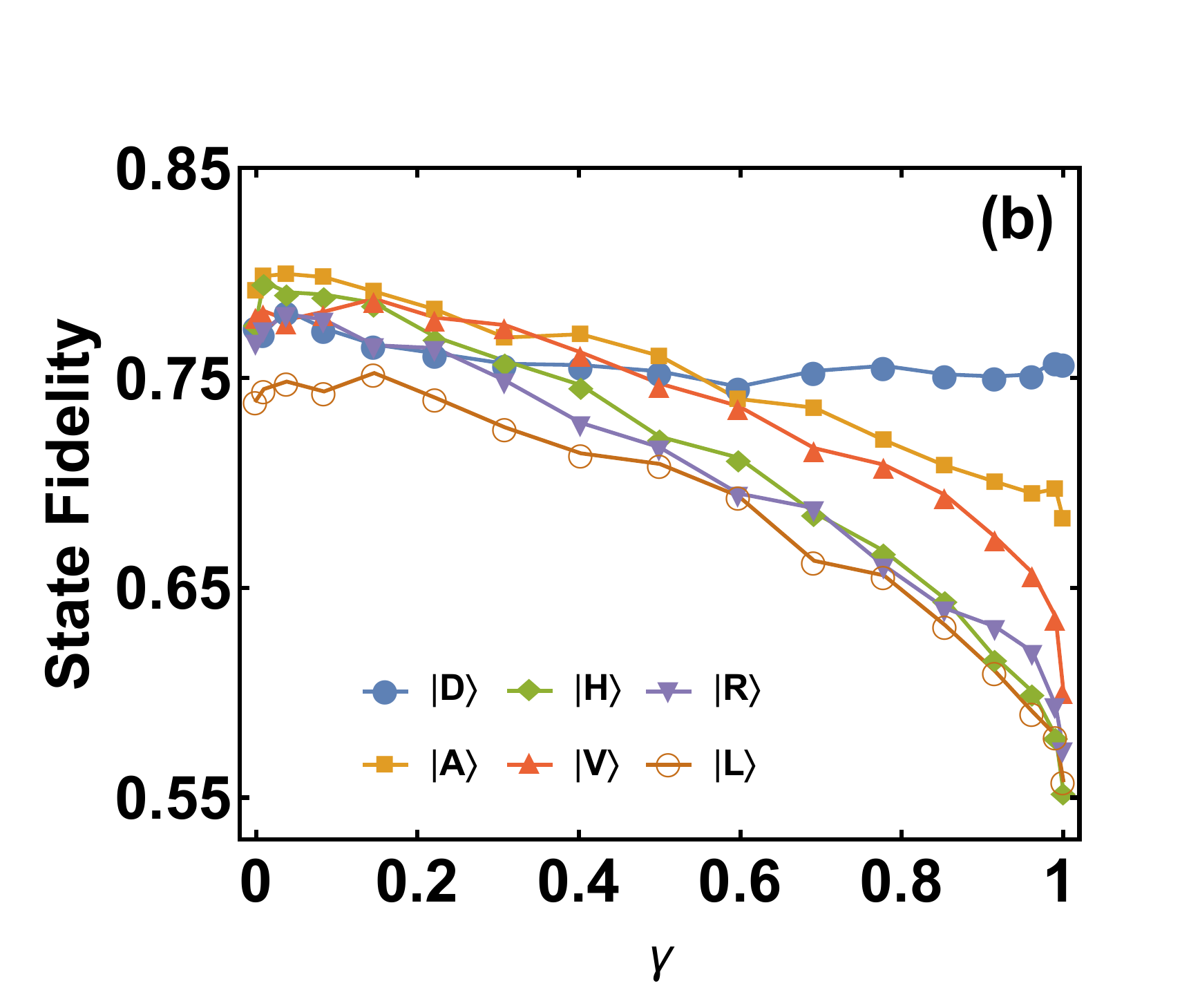}\vskip-5mm
		\caption*{(b) Optimal Recovery}
		\label{figS2:side:b}
	\end{minipage}
	\begin{minipage}[t]{0.52\linewidth}
		\centering \vskip-1cm
		\includegraphics[width=3.8in]{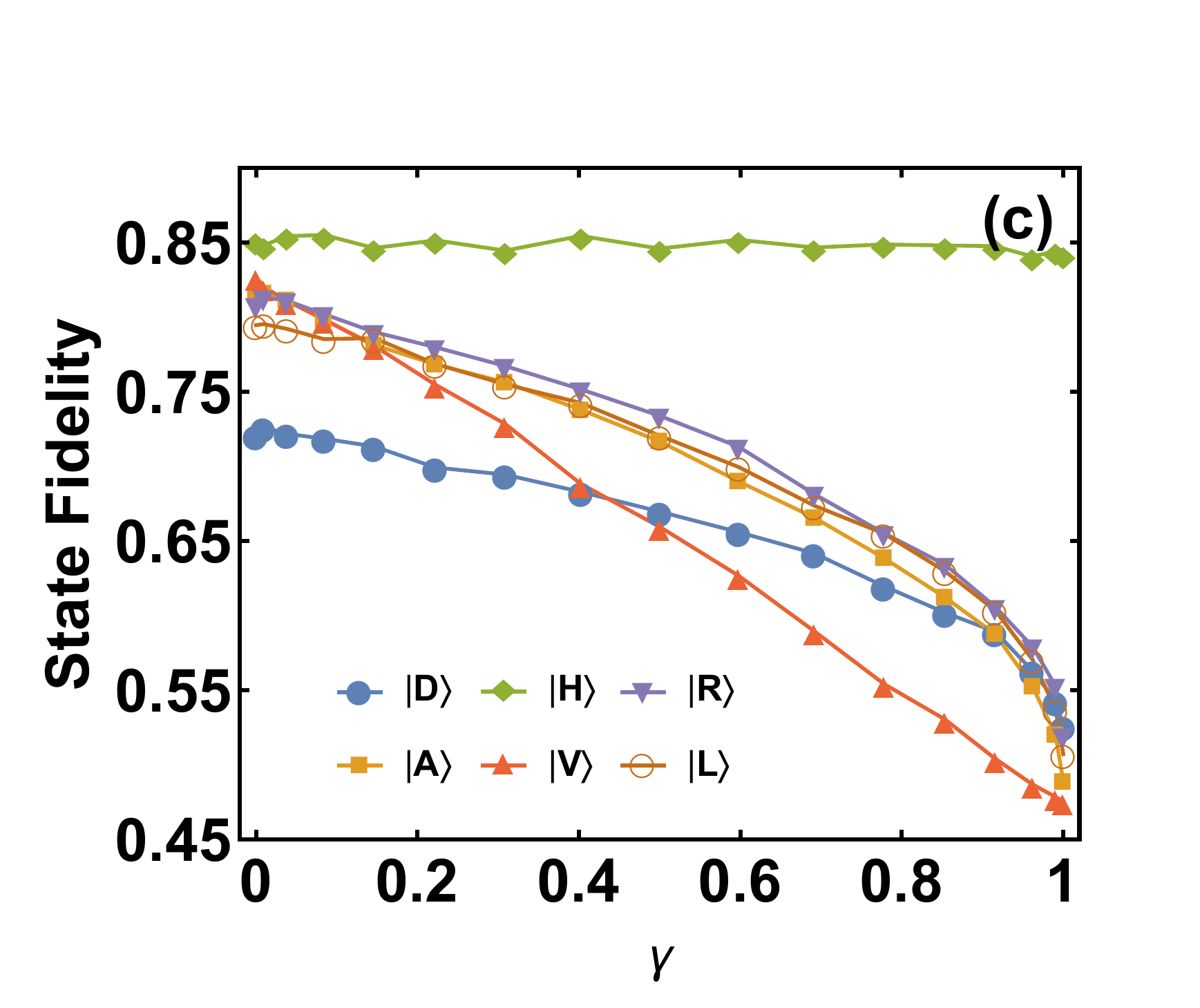}\vskip-5mm
		\caption*{(c) Without error correction}
		\label{figS2:side:c}
	\end{minipage}
	\caption{\label{figS2}Experimental results from IBM Q. Optimal
          Recovery shows a dominance at large damping ratio
          $\gamma$. The error bars in Fig.~\ref{Fig.result}~(a)
          are derived from the standard deviation via bootstrapping. }
\end{figure}
\begin{figure}[H]
	\begin{minipage}[t]{1\linewidth}
		\centering \vskip-1cm
		\includegraphics[width=4.3in]{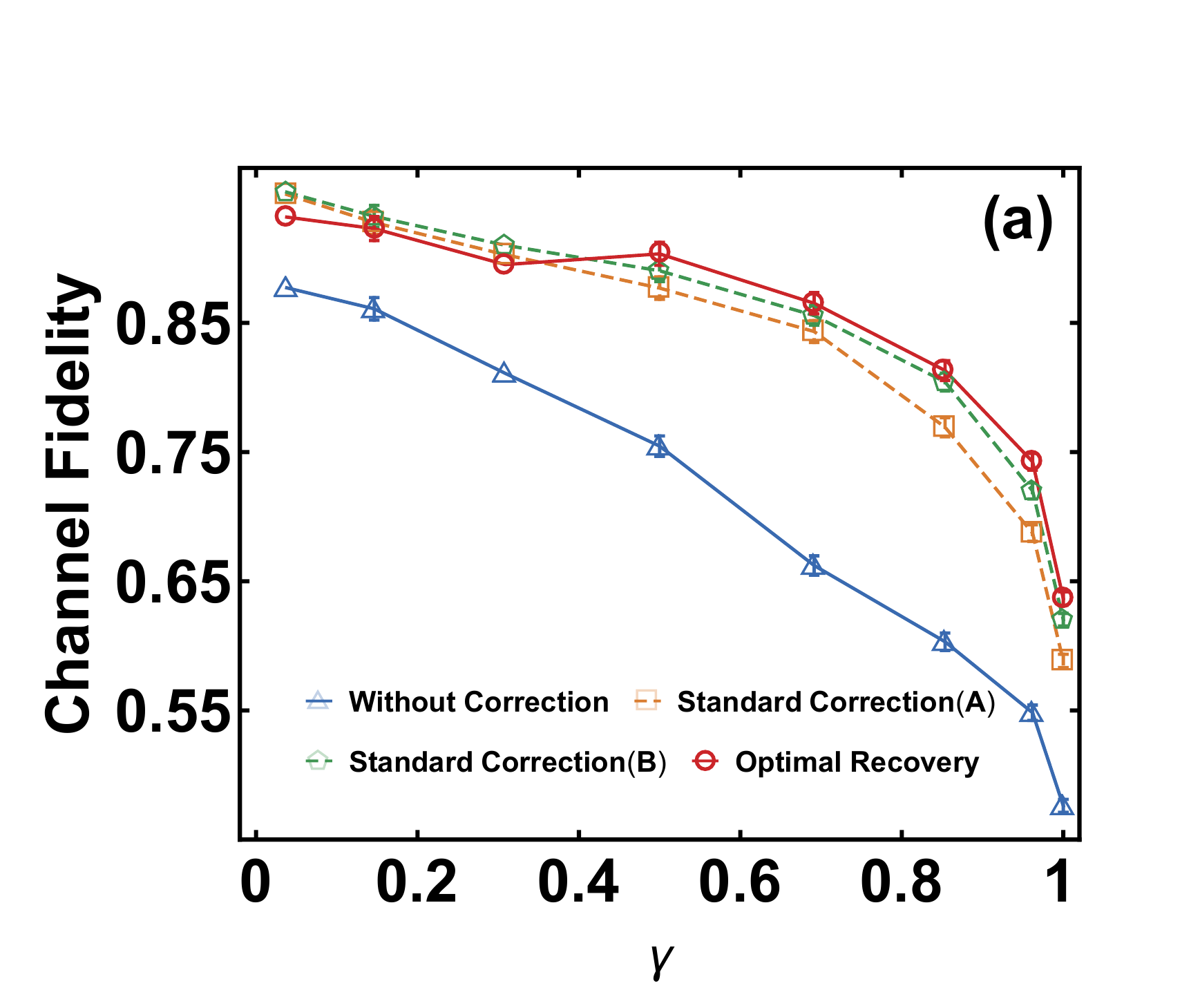}\vskip-5mm
		\caption*{(a) Channel fidelities}
		\label{figS3:side:a}
	\end{minipage}
	\begin{minipage}[t]{0.5\linewidth}
		\centering \vskip-1cm
		\includegraphics[width=3.8in]{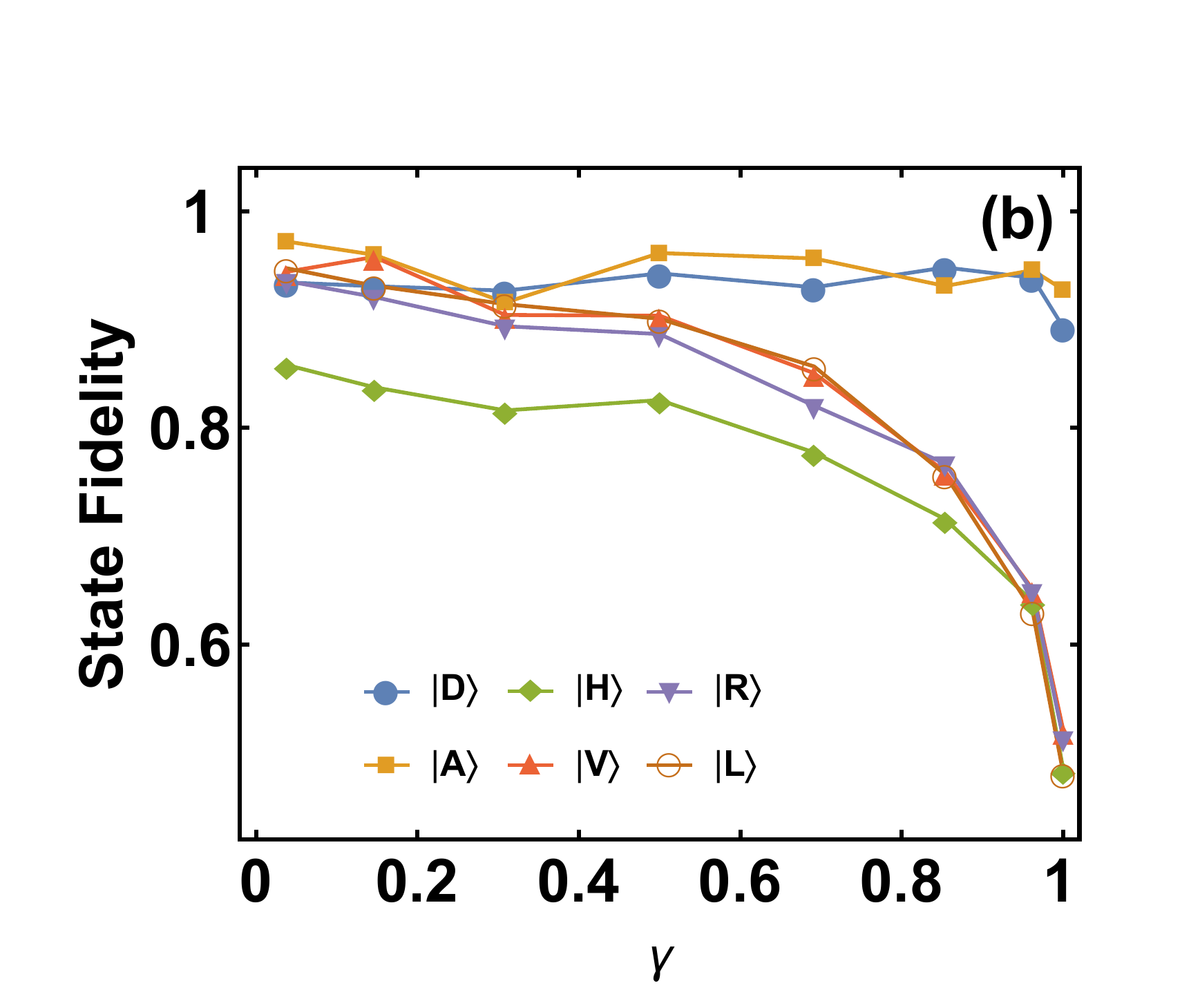}\vskip-5mm
		\caption*{(b) Optimal Recovery}
		\label{figS3:side:b}
	\end{minipage}
	\begin{minipage}[t]{0.5\linewidth}
		\centering \vskip-1cm
		\includegraphics[width=3.8in]{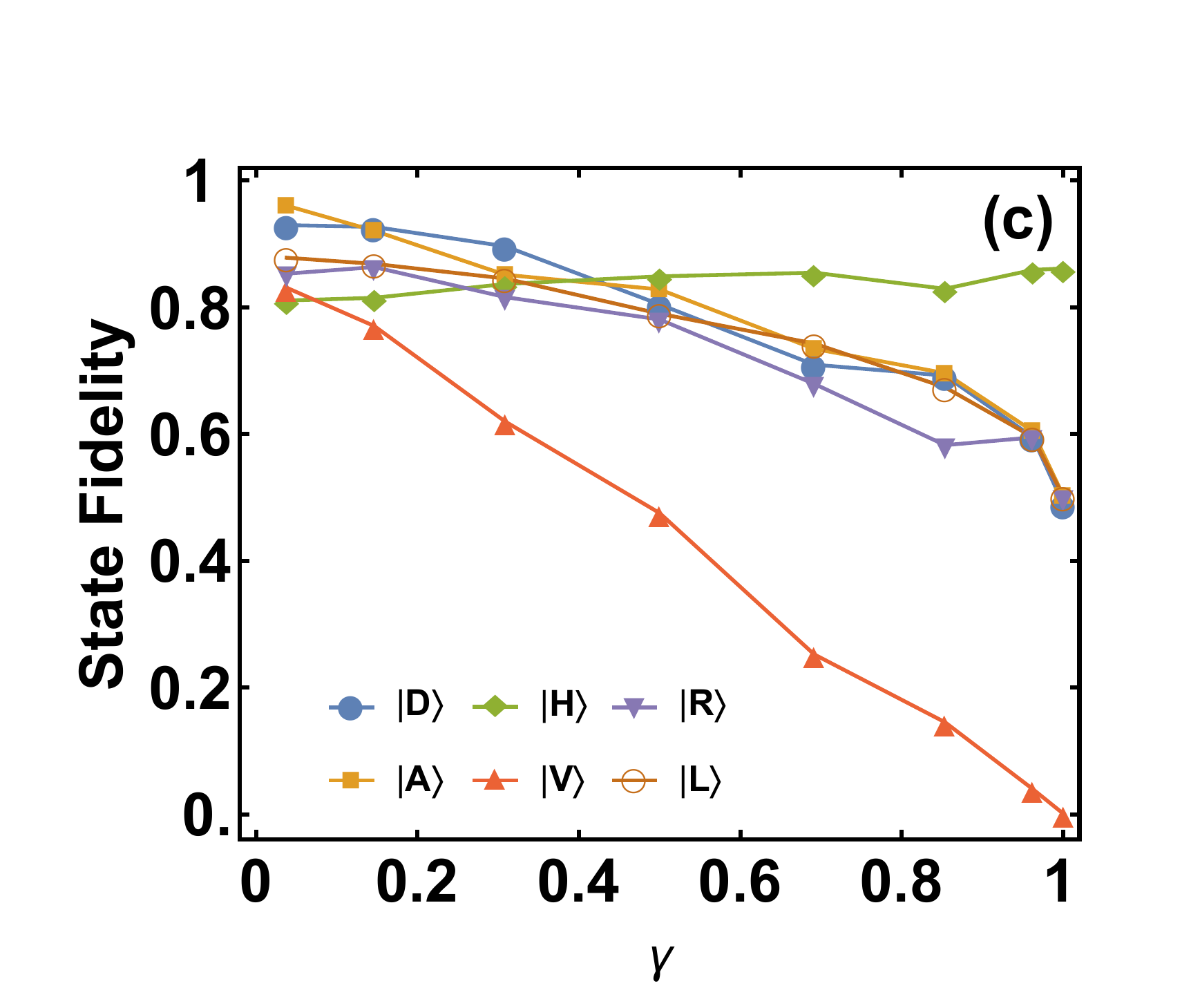}\vskip-5mm
		\caption*{(c) Without Correction}
		\label{figS3:side:c}
	\end{minipage}
	\caption{\label{figS3}Experimental results from the NMR
          system. The experiment on the NMR system verifies the power of
          Optimal Recovery. The error bars in diagram (a) are from random samples.}
\end{figure}
\begin{figure}[H]
	\begin{minipage}[t]{0.5\linewidth}
		\centering \vskip-1cm
		\includegraphics[width=3.7in]{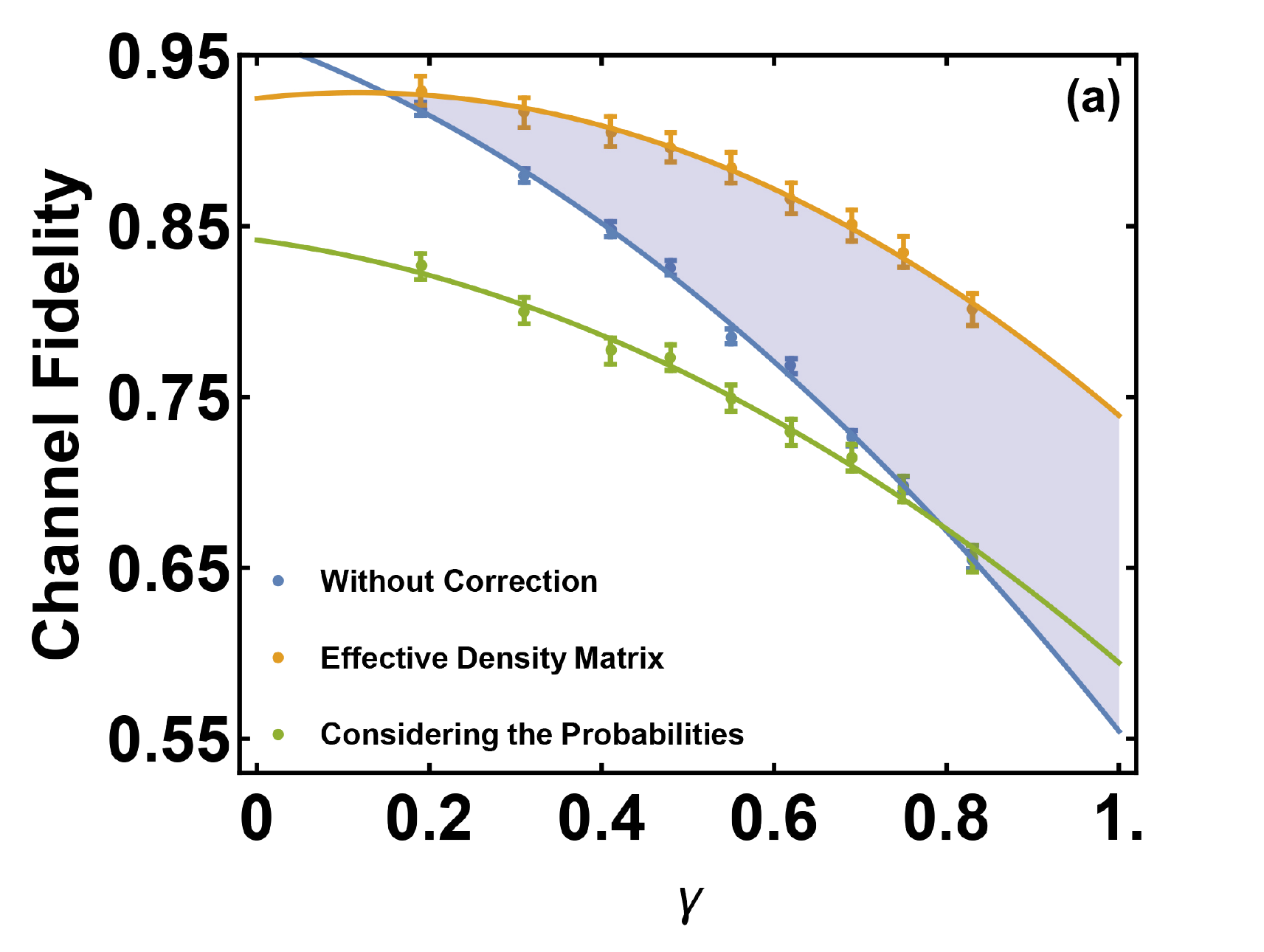} \vskip-5mm
		\caption*{(b) Shot loss on the optical platform}
		\label{figS4:side:a}
	\end{minipage}
	\begin{minipage}[t]{0.5\linewidth}
		\centering \vskip-1cm
		\includegraphics[width=3.7in]{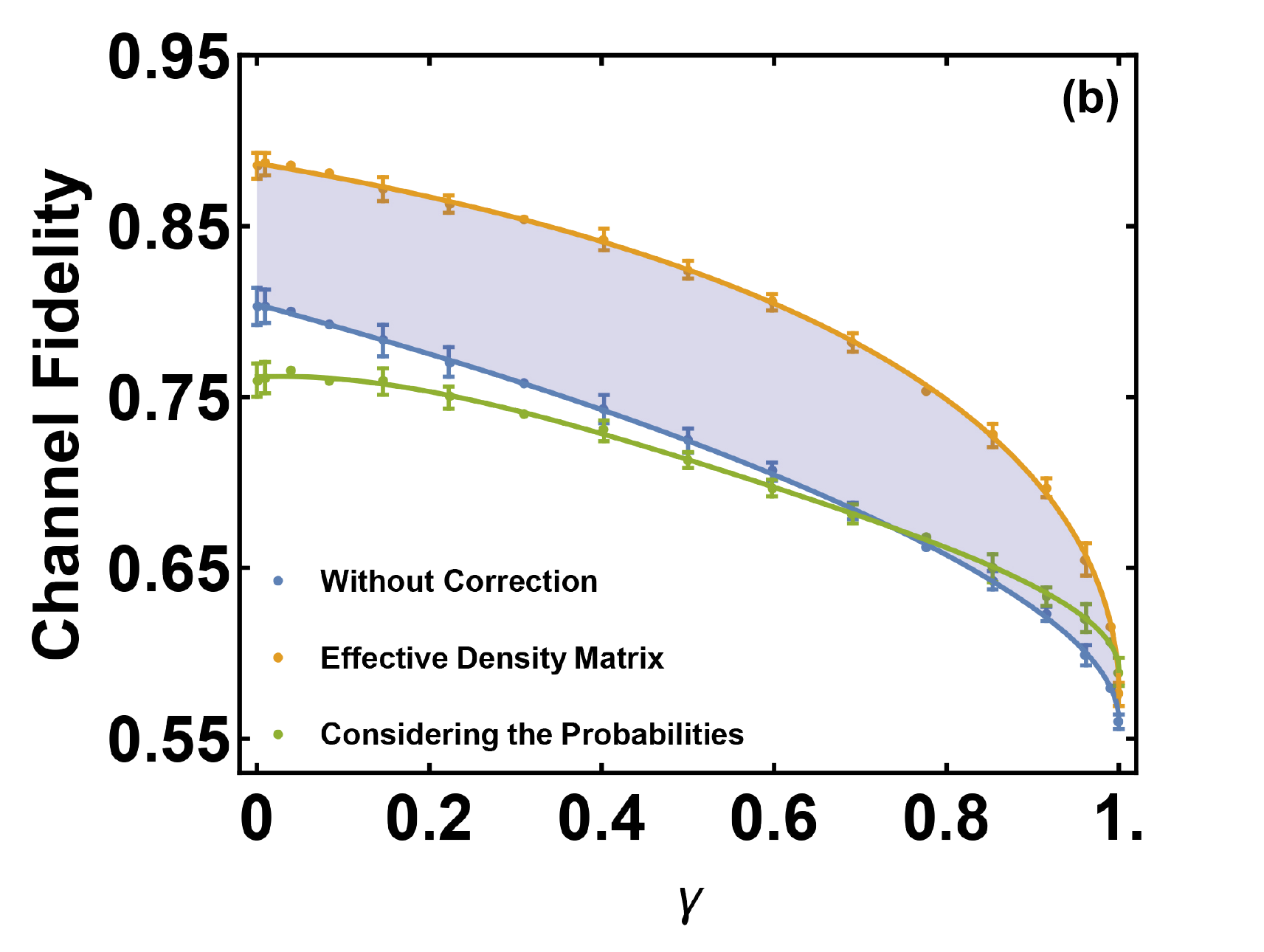} \vskip-5mm
		\caption*{(a) Shot loss on IBM Q}
		\label{figS4:side:b}
	\end{minipage}
	\caption{\label{figS4}Shot loss of CNOT. (a) On the optical
          platform, the partially polarizing beam splitters (PPBS)
          mainly contribute to the loss of photons. Compared with the
          raw single-qubit density matrix, the fidelity of the
          density matrix incorporating the loss declines a lot. (b) On
          IBM Q, the loss of shots results from the crosstalk between
          qubits, and leads to some abnormal excitation on other
          qubits.  We compare the fidelity of the raw single-qubit
          density matrix with that of the real density matrix
          considering the probabilities.}
\end{figure}

\end{document}